\documentclass[apj]{emulateapj}
\slugcomment{{\sc Accepted to ApJ:} January 30, 2009} 
\pdfoutput = 1

\usepackage{natbib}
\usepackage{graphicx}

\providecommand{\e}[1]{\ensuremath{\times 10^{#1}}}

\shorttitle{Properties of Dense Cores}
\shortauthors{Foster et al.}
\begin{document}
\title{Dense Cores in Perseus: The Influence of Stellar Content and Cluster Environment}

\author{Jonathan B. Foster\altaffilmark{1}, Erik Rosolowsky\altaffilmark{2}, Jens Kauffmann\altaffilmark{1,3}, Jaime Pineda\altaffilmark{1}, Michelle Borkin\altaffilmark{3}, Paola Caselli\altaffilmark{4}, Phil Myers\altaffilmark{1}, Alyssa Goodman\altaffilmark{1,3}}
\altaffiltext{1}{Harvard-Smithsonian Center for Astrophysics, 60 Garden Street, Cambridge, MA 02138}
\altaffiltext{2}{University of British Columbia Okanagan, 3333 University Way, Kelowna, BC V1V 1V7, Canada}
\altaffiltext{3}{Initiative in Innovative Computing, Harvard
  University, 60 Oxford St., Cambridge, MA 02138}

\altaffiltext{4}{School of Physics and Astronomy,
University of Leeds, Leeds LS2 9JT, UK}

\begin{abstract}

We present the chemistry, temperature, and dynamical state of a sample of 193 dense cores or core candidates in the Perseus Molecular cloud and compare the properties of cores associated with young stars and clusters with those which are not. The combination of our NH$_3$ and CCS observations with previous millimeter, sub-millimeter, and Spitzer data available for this cloud enable us both to determine core properties precisely and to accurately classify cores as starless or protostellar. The properties of cores in different cluster environments and before-and-after star formation provide important constraints on simulations of star-formation, particularly under the paradigm that the essence of star formation is set by the turbulent formation of prestellar cores. We separate the influence of stellar content from that of cluster environment and find that cores within clusters have (1) higher kinetic temperatures (12.9 K vs. 10.8 K) and (2) lower fractional abundances of CCS (0.6\e{-9} vs 2.0\e{-9}) and NH$_3$ (1.2\e{-8} vs 2.9\e{-8}). Cores associated with protostars have (1) slightly higher kinetic temperatures (11.9 K vs. 10.6 K) (2) higher NH$_3$ excitation temperatures (7.4 K vs. 6.1 K), (3) are at higher column density (1.2\e{22} cm$^{-2}$ vs. 0.6\e{22} cm$^{-2}$), have (4) slightly more non-thermal/turbulent NH$_3$ linewidths (0.14 km/s vs. 0.11 km/s FWHM), have (5) higher masses (1.5 M$_{\odot}$ vs. 1.0 M$_{\odot}$) and have (6) lower fractional abundance of CCS (1.4\e{-9} vs. 2.4\e{-9}). All values are medians. We find that neither cluster environment nor protostellar content makes a significant difference to the dynamical state of cores as estimated by the virial parameter -- most cores in each category are gravitationally bound. Only the high precision of our measurements and the size of our sample makes such distinctions possible. Overall, cluster environment and protostellar content have a smaller influence on the properties of the cores than is typically assumed, and the variation within categories is larger than the differences between categories.

\end{abstract}


\section{Introduction}
Stars form in dense cores, providing ample motivation for understanding these enigmatic objects. Several authors have claimed that the stellar initial mass function (IMF) comes directly from the core-mass function (CMF) with a Saltpeter power-law slope at high masses (dN(m)/d $log (m) \sim m^{-1.35}$) and a turnover at some characteristic mass, although the value of this mass is in dispute (compare \citet{Motte:1998} with \citet{Alves:2007}). This correspondence between the mass functions suggests two important corollaries: that star formation is a simple process which transforms core mass into stellar material in a robust and largely invariant fashion; and that a theory of molecular cloud formation and evolution which provides the correct core-mass function (e.g. the turbulent fragmentation model of \citet{Padoan:2002}) has solved a critical part of the star formation problem. However, \citet{Swift:2008} model several diverse evolutionary schemes which map a CMF onto an IMF and show that the present data on mass functions is insufficient to distinguish between their different schemes. Other properties of the dense core population are required to constrain theoretical models.

The definition of a dense core differs depending on the method employed, sensitivity achieved, and physical resolution reached. The platonic ideal is a small (0.1 pc),  cold ($\sim$ 10K), dense ($> 10^{4}$ cm$^{-3}$), quiescent (small non-thermal linewidth) object \citep{Myers:1983,Benson:1989,Goodman:1998}, with a strong density contrast from its surroundings and no significant substructure, allowing a simple radial power-law description of their structure \citep[e.g.][]{Ward-Thompson:1994}. In addition, if we are searching for the precursor to stars we want to exclude transient objects; a dense core should be close to gravitationally bound. However, gravitationally unbound cores may be compressed by turbulence and thus still collapse \citep{Gomez:2007}. It is also not clear whether we should exclude a dense core which contains a young star. We know that such an object is the formation site of a star, but the protostar may have had significant impact upon the core, rendering it useless for studying the initial conditions of star formation. 

Many detailed studies have been carried out on individual dense cores (see \citet{Bergin:2007} for a review). Much can also be gained from cloud-wide studies of cores, for though they must sacrifice some level of detail, such studies are able to average over certain inevitable biases (such as lacking velocity information on the plane of the sky and spatial information along the line of sight), provide good statistics of the population as a whole, and allow us to assess the influence of environmental variations. One particularly powerful combination is that of NH$_3$ observations with either continuum emission (typically millimeter or sub-millimeter) or extinction maps which trace the dust within cores. Sufficiently high signal-to-noise ammonia spectra provide a measurement of the temperature and the total linewidth of an object, so we are able to say whether an object is cold and quiescent. The critical density of ammonia is roughly 10$^{4}$ cm$^{-3}$ \citep{Ho:1983}, so objects detected in NH$_3$ are typically at or above this density and thus more dense than the bulk of the molecular cloud. Ammonia also provides one piece of the stability puzzle, namely how much internal non-thermal velocity dispersion other forces must overcome in order to collapse the core. Millimeter maps can provide spatial extent, disambiguation from the background, and masses (with the aid of measured or assumed temperatures). This provides a first order estimate of whether a core is bound, though magnetic support and external pressure still elude us. Stellar content can be assessed through direct infrared observations of point sources with an instrument like Spitzer.

We present such a synthesis of observations for the molecular cloud in Perseus by combining the results of an ammonia survey of core-candidates with the Green Bank Telescope \citep{Rosolowsky:2008} with a Bolocam survey of the continuum at 1 mm \citep{Enoch:2006} and a catalog of young stellar objects from the Cores-to-Disks Spitzer Legacy project \citep{Evans:2003}. These three data sets are all cloud-wide and spatially complete. The NH$_3$ survey covers every dense core and dense core candidate within the cloud, drawn principally from the Bolocam survey with additional candidates drawn from additional COMPLETE\footnote[1]{COordinated Molecular Probe Line Extinction Thermal Emission Survey of Star Forming Regions} \citep{Ridge:2006} data as described in \citet{Rosolowsky:2008}. 

Our analysis includes fewer cores (of the 193 pointings, 122 are Bolocam cores) than the seminal work of \citet{Jijina:1999} (264 cores), but is complete for Perseus down to the mass limit of the Bolocam survey, and is more uniform. This uniformity comes both from using a consistent observational setup and from the fact that we are studying one cloud at a fixed distance. 

The study of the Pipe Nebula in \citet{Lada:2008} is a comparable cloud-wide study based on extinction map made from 2MASS, C$^{18}$O observations from the Arizona Radio Observatory 12m \citep{Muench:2007}, and NH$_3$ observations from the GBT \citep{Rathborne:2008}. Because Perseus has more massive cores and is at a better declination for observing from the GBT, we are able to detect NH$_3$ from every Bolocam core. This allows us to measure the temperature of each core. The Pipe also has extremely limited star formation, and may represent an early stage of star formation, while Perseus provides examples of both isolated and clustered star formation as well as relatively pristine starless objects.

This paper uses a statistical approach to separate and measure the influence of both stellar content and cluster environment on dense cores. Our main focus will be on the chemistry, temperature, and dynamical states of the cores. The mass distributions of cores throughout Perseus have already been explored in detail by a number of papers using two thermal emission surveys: the Bolocam \citep{Enoch:2006} and SCUBA \citep{Hatchell:2005} surveys of Perseus. Follow-up studies have examined the distinction between mass distributions of protostellar and starless cores \citep{Hatchell:2008,Enoch:2008,Kirk:2006}, so we will touch but lightly on this question. The stability or dynamical state of these cores have also been previously studied \citep{Kirk:2007,Enoch:2008}, but our data sets allow for a more detailed analysis. 

The distinction between clustered and isolated cores was made in \citet{Jijina:1999}, but has not since received sufficient attention; our cloud-wide survey is the perfect data-set for such a study, as Perseus features two main clusters as well as star formation in isolated regions. Recent models of star formation \citep[e.g.][]{Nakamura:2007} which include the role of feedback clearly establish the need to assess the influence of environment upon the star formation process, an influence which may manifest itself in the core properties. If cores within clusters are substantially different from isolated cores, this may explain why the CMF appears to be different in different regions. It is also possible that the role of feedback means that star formation within clusters is a different process than star formation in isolated regions and that the notion of a simple function which maps the CMF to the IMF must at least be modified to allow for different environments. 

We describe our datasets in \S\ref{Datasets} and describe our statistical tests in \S\ref{Statistical Methods}. In \S\ref{Results} we present in full detail the distributions of the following variables: kinetic temperature, excitation temperature, column density, fractional abundance of NH$_3$ and CCS, linewidths (total and non-thermal), mass, and stability (the virial parameter, $\alpha$). Along the way, we compare with results obtained for Perseus in previous studies, and present a comparison with studies in other regions in \S\ref{Comparison}. Distributions of all variables by category are provided in Table~\ref{SummaryOfDist}, while a summary of which variables are significantly influenced by cluster environment or protostellar content is provided in Table~\ref{ANOVAtable}.

\section{Description of Datasets}
\label{Datasets}

\subsection{Ammonia Survey}
\label{nh3survey}
The ammonia data in this paper are taken from the spectral atlas described in \citet{Rosolowsky:2008}. The survey consists of 193 spectra taken with the Green Bank Telescope, pointed towards a selection of cores and core candidates in the Perseus molecular cloud.  The targets were drawn from a merged list designed for maximum completeness based on several surveys.  In addition to cataloged sources of millimeter continuum emission, the survey targeted marginal millimeter sources and compact regions of extinction.  The GBT spectrometer was configured to produce high resolution (0.025 km~s$^{-1}$) spectra of NH$_3$ (1,1) and (2,2) emission as well as CCS ($2_1\to 1_0$) and CC$^{34}$S ($2_1\to 1_0$).  The beam size of the GBT is 31\arcsec at these frequencies which projects to a size of 0.04 pc at the distance of Perseus \citep[250 pc, see][for a full discussion of the distance]{Enoch:2006}.

Following reduction, a uniform slab radiative transfer solution was fit to the full ammonia complex to determine the values of the kinetic temperature ($T_\mathrm{kin}$), the line-of-sight velocity ($v_\mathrm{lsr}$), the velocity dispersion of the tracer ($\sigma_\mathrm{NH3}$), the radiative excitation temperature ($T_\mathrm{ex}$) and the full line optical depth ($\tau$).  Using a combination of the derived parameters, the column densities of both NH$_3$ and CCS were also derived.  Since the model is necessarily simple, complicated sources will be represented by average properties. However, the model produces good fits for the wide variety of sources found in the survey data. In cases where the spectra were clearly separable into two distinct components such fitting was performed, but these results were flagged and excluded from our analysis of variables which rely upon an integrated column measurement (fractional abundances, dust masses, and the virial parameter).

\subsection{Bolocam Survey} 

For a fiducial sample of dense cores, we used the millimeter continuum observations of dust emission presented by \citet{Enoch:2006}. That work used Bolocam, a bolometer array on the Caltech Submillimeter Observatory, to map the high extinction portion ($A_V\gtrsim 2$) of the Perseus molecular cloud at $\lambda\approx 1.1$~mm.  Using typical assumptions about dust emission, the Bolocam survey is complete down to $\sim 0.2~M_{\odot}$ at the assumed distance of Perseus.  However, owing to variations in the degree to which sources are recovered under the Bolocam mapping strategy, the mass sensitivity is governed by the source size.  In addition, large scale structure is filtered out by the adopted observation strategy, though this filtering tends to highlight the dense cores desired for study.

Dense cores are identified by convolving the final map by an optimal filter that selects point sources.  From a list of significant peaks in the filtered map, only sources with a well defined centroid are considered dense cores.  Fluxes are measured in apertures around each object and major/minor axis sizes are from two-dimensional Gaussian fits to the emission profile.  A total of 122 dense cores are so identified and cataloged.  These cores were the primary catalog adopted for the spectral line survey (\S\ref{nh3survey}) owing to the complete spatial coverage provided by the Bolocam survey and the similarity in resolution to the GBT (also $31''$).  The Bolocam survey likely identifies every region harboring dense cores, though higher resolution observations \citep{Kirk:2006} suggest that some of the cores may have significant substructure.

Although there is evidence that in certain high-density and high-radiation environments NH$_3$ does a poor job of tracing some dust cores (see the study of \citet{Friesen:2008} in Ophiuchus), maps of regions of Perseus made with GBT NH$_3$ observations similar to those in this pointed survey show an excellent correspondence with dust maps such as the BOLOCAM map used in this study \citep{Pineda:2009}.

\subsection{YSO and Cluster identification}
\label{ID}

\begin{deluxetable}{lcc}
\tablewidth{0pc}
\tablecaption{Number of cores in each category}
\tablehead{ \colhead{} & \colhead{Starless} & \colhead{Protostellar} \\
\colhead{} & \colhead{certain(uncertain)} & \colhead{certain(uncertain)}}
\startdata
Clustered & 23 (7) & 16 (3) \\
Isolated & 102 (4) & 38 (6) \\
\cutinhead{Subset of cores with NH$_3$ detections}
Clustered & 19 (7) & 15 (3)\\
Isolated & 79 (4)  & 35 (6) \\
\enddata
\label{Numbers}
\end{deluxetable}

Classifying cores as starless or protostellar is a difficult problem. In the case of Perseus, several previous studies have examined this issue from different perspectives \citep{Hatchell:2007,Enoch:2008}, so our approach is to synthesize these efforts with some of our own methods. The areas of disagreement between these methods hint at the complexity of this problem which is exacerbated by differing definitions as well as observational uncertainty. Full details of this synthesis are presented in Appendix A, but we take the abundantly cautious approach of tagging each core as either (1) Starless, (2) Probably Starless, (3) Probably Protostellar, or (4) Protostellar, with the uncertain classifications resulting from disagreements between the various methods used. This classification allows us to verify that cores with uncertain classification do not influence our results. Our final list of starless/protostellar for these NH$_3$ pointings is presented in Table \ref{IDs} in Appendix A, with comments.

We classified each core as isolated or clustered based on the definitions of the cluster boundaries of IC 348 and NGC 1333 in \citet{Jorgensen:2006}, identified as high concentrations of YSO candidates from Spitzer data. For NGC1333 this means from 03:28:00 to 03:30:00 in RA and +31:06:00 to +31:30:00 in Dec, while IC 348 goes from 03:43:12 to 03:46:00 in RA and +31:48:00 to +32:24:00 in Dec. For NGC 1333 there is a clear drop of in the density of objects beyond this boundary, while in IC 348 the cut-off at the western edge is slightly more ambiguous, but different definitions of the clusters would change only a small number of objects.

The number of cores within each category is presented in Table \ref{Numbers}. These numbers represent the upper limits on how many objects may be considered in our analyses. For each variable, we use only objects for which that variable is well measured. In particular, some objects lack either significant NH$_3$ or millimeter emission and are unlikely to be genuine cores.  

If we restrict ourselves to objects with a NH$_3$ detection, the ratio of starless to protostellar cores is higher in isolated regions (N$_\mathrm{starless}$/N$_\mathrm{protostellar}$ = 2.0) than in clustered regions  (N$_\mathrm{starless}$/N$_\mathrm{protostellar}$ = 1.4). Both these ratios are higher than found in \citet{Enoch:2008}, where (N$_\mathrm{starless}$/N$_\mathrm{embedded}$ = 1.0) for Perseus. This is partly due to our different classification of which cores are starless as explained in Appendix A, but is mostly due to the extra 71 positions we target which are not in the Bolocam sample and are largely starless. The diverse datasets from which these additional positions were drawn prohibit us from making a strong inference from these ratios.

\section{Statistical Methods}
\label{Statistical Methods}

\subsection{Distributions of Variables}
\label{DistSection}
The following variables are derived as part of the fit of the NH$_3$ and CCS spectra in \citet{Rosolowsky:2008}: T$_\mathrm{kin}$, T$_\mathrm{ex}$, NH$_3$ and CCS column density, NH$_3$ and CCS total linewidths. In this paper, we use T$_\mathrm{kin}$ from the NH$_3$ line and Bolocam fluxes derived in \citet{Enoch:2006} in order to derive total column densities (\S\ref{ColDensity}) and fractional abundances of NH$_3$ and CCS (\S\ref{Chemistry}). We use T$_\mathrm{kin}$ and the total linewidths to calculate non-thermal linewidth  of both NH$_3$ and CCS (\S\ref{Line-widths}). Finally, we use T$_\mathrm{kin}$ to improve the estimated masses (\S\ref{MassStability}) of Bolocam cores derived in \citet{Enoch:2006} and from this and the non-thermal linewidths we estimate the virial parameter (\S\ref{MassStability}) for each core. The derivation of new quantities is presented in the sections referenced above.

Table \ref{SummaryOfDist} presents a summary of the distributions of all our variables, broken down by classification. Values are presented for the mean, standard deviation, median, and 25\% and 75\% quartiles, as many variables are not well described as a normal distribution. As we analyze only cores where a given variable can be well determined, we suffer some bias due to non-detections of small values. This is possibly a concern in T$_\mathrm{kin}$ and is definitely a concern for the fractional abundance of CCS (X-CCS), mass, and the virial parameter, $\alpha$. We discuss these biases within the text for each variable. 

Overall, the variation within categories is generally larger than the differences between categories and only the high precision of our measurements and the size of our sample makes it possible to draw distinctions between categories. Table~\ref{ANOVAtable} highlights the distributions where statistically significant differences exist, as determined by an Analysis of Variance test, described below.

\begin{deluxetable*}{llccccc}
\tablecolumns{7}
\tabletypesize{\scriptsize}
\tablewidth{0pc}
\tablecaption{Summary of Variable Distributions by Classification}
\tablehead{Variable & Classification\tablenotemark{a} & Mean & $\sigma$ & Median & 25\% quartile & 75\% quartile}
\startdata
T$_\mathrm{kin}$ (K) & Protostellar & 12.67(12.51) & 2.4(2.3) & 11.85(11.83) & 11.32(11.01) & 14.05(13.49) \\
Kinetic temperature & Starless & 11.43(11.75) & 2.7(3.1) & 10.55(10.67) & 9.99(10.02) & 11.66(12.41) \\
 & Cluster & 13.35 & 3.2 & 12.92 & 11.15 & 14.40 \\
 & Isolated & 11.51 & 2.5 & 10.79 & 10.08 & 11.96 \\
\\
T$_\mathrm{ex}$ (K) & Protostellar & 7.25(7.12) & 1.2(1.3) & 7.35(7.26) & 6.46(6.19) & 8.32(8.08) \\
Excitation temperature & Starless & 5.99(6.13) & 0.8(0.9) & 6.09(6.13) & 5.58(5.74) & 6.57(6.60) \\
 & Cluster & 6.82 & 1.3 & 6.57 & 6.03 & 7.82 \\
 & Isolated & 6.48 & 1.2 & 6.41 & 5.83 & 7.35 \\
\\
N(H$_2$) (1\e{22} cm$^{-2}$) & Protostellar & 2.22(2.08) & 2.7(2.5) & 1.18(1.18) & 0.70(0.71) & 2.94(2.59) \\
Total column density & Starless & 0.66(0.73) & 0.5(0.6) & 0.59(0.63) & 0.28(0.28) & 0.85(0.93) \\
 & Cluster & 1.13 & 1.0 & 0.93 & 0.52 & 1.42 \\
 & Isolated & 1.23 & 1.9 & 0.70 & 0.44 & 1.24 \\
\\
X-NH$_3$ (1\e{-8}) & Protostellar & 5.39(5.15) & 13.5(12.5) & 3.39(3.39) & 1.06(1.09) & 5.00(5.04) \\
Fractional abundance of NH$_3$ & Starless & 8.17(7.61) & 32.3(31.0) & 2.31(1.87) & 0.63(0.52) & 4.40(4.16) \\
 & Cluster & 1.62 & 1.4 & 1.18 & 0.52 & 2.59 \\
 & Isolated & 8.57 & 30.8 & 2.89 & 0.86 & 5.19 \\
\\
X-CCS (1\e{-9}) & Protostellar & 1.48(1.83) & 0.9(2.1) & 1.43(1.43) & 0.79(0.79) & 2.06(2.06) \\
Fractional abundance of CCS & Starless & 3.24(3.11) & 2.5(2.5) & 2.37(2.32) & 1.70(1.58) & 4.65(4.32) \\
 & Cluster & 0.86 & 0.4 & 0.60 & 0.55 & 1.50 \\
 & Isolated & 2.87 & 2.5 & 1.96 & 1.43 & 3.24 \\
\\
$\sigma_\mathrm{non-thermal}$ (NH$_3$) (km/s) & Protostellar & 0.142(0.138) & 0.04(0.04) & 0.135(0.134) & 0.113(0.113) & 0.176(0.174) \\
Non-thermal linewidth & Starless & 0.117(0.122) & 0.04(0.04) & 0.114(0.117) & 0.097(0.099) & 0.137(0.148) \\
 & Cluster & 0.138 & 0.04 & 0.135 & 0.114 & 0.169 \\
 & Isolated & 0.124 & 0.04 & 0.122 & 0.103 & 0.145 \\
\\
$\sigma_\mathrm{non-thermal}$ (CCS) (km/s) & Protostellar & 0.196(0.195) & 0.09(0.08) & 0.181(0.181) & 0.153(0.153) & 0.226(0.226) \\
Non-thermal linewidth  & Starless & 0.165(0.172) & 0.08(0.08) & 0.151(0.155) & 0.120(0.120) & 0.180(0.210) \\
 & Cluster & 0.135 & 0.06 & 0.129 & 0.129 & 0.155 \\
 & Isolated & 0.186 & 0.08 & 0.165 & 0.144 & 0.217 \\
\\
Mass (M$_{\odot}$) & Protostellar & 2.37(2.44) & 2.8(3.0) & 1.50(1.46) & 0.76(0.76) & 2.80(3.45) \\
M$_\mathrm{dust}$& Starless & 1.41(1.35) & 0.9(0.9) & 1.02(1.02) & 0.82(0.82) & 1.90(1.81) \\
 & Cluster & 1.57 & 1.3 & 1.14 & 0.76 & 1.92 \\
 & Isolated & 1.91 & 2.3 & 1.20 & 0.81 & 2.30 \\
\\
Virial parameter: $\alpha$ & Protostellar & 1.36(1.31) & 0.8(0.8) & 0.99(0.99) & 0.80(0.75) & 1.97(1.97) \\
M$_\mathrm{dynamical}$/M$_\mathrm{dust}$  & Starless & 1.37(1.35) & 0.5(0.5) & 1.39(1.39) & 1.12(1.04) & 1.64(1.58) \\
 & Cluster & 1.47 & 0.8 & 1.37 & 0.82 & 2.06 \\
 & Isolated & 1.28 & 0.6 & 1.24 & 0.84 & 1.57 \\
\\
\label{SummaryOfDist}
\enddata
\tablenotetext{a}{For the protostellar/starless classifications, the results are displayed as: certain(including uncertain).}
\end{deluxetable*}

\subsection{Analysis of Variance Statistical Test}
\label{ANOVASection}

In order to quantify the respective influence of stellar content and cluster environment, we performed Analysis of Variance tests (ANOVA) on a number of key parameters. ANOVA tests \citep[see][for a complete reference]{Scheffe:1959} are useful for assessing the influence of categorical or qualitative variables on other variables. In this case, we categorize each core as starless/protostellar and clustered/isolated and test the hypothesis that these categories have a significant influence on the measured quantitative variables. In particular, the Two-Way ANOVA framework allows us to construct a statistical model to test for the ``interaction effect'' by comparing a model containing an interaction term:
\begin{equation}
Y_{ijk} = \mu + \alpha_{i} + \beta_{j} + (\alpha \beta)_{ij} + \epsilon_{ijk},
\end{equation}
against a model without this term,
\begin{equation}
Y_{ijk} = \mu + \alpha_{i} + \beta_{j} + \epsilon_{ijk},
\end{equation}
via the F-test. In these models we have two variables, $\alpha$ at i levels and $\beta$ at j levels. We test the hypothesis that each individual core's property (Y$_{k}$) is due to some mean value ($\mu$), some contribution from protostellar content ($\alpha$), some from environment ($\beta$), some from the interaction effect ($\alpha \beta$), and some from random error ($\epsilon$). 

A simple hypothetical example of the interaction effect would be the following: starless cores are 2 K colder than protostellar cores, and isolated cores are 1 K colder than clustered cores, but starless AND isolated cores are 5 K colder than protostellar cores in clusters -- i.e. the influence of the two variables is not simply additive/linear. In our study it turned out that in all cases the interaction effect was statistically insignificant, which allows us to make direct statements about the influence of the individual categories (starless/protostellar and clustered/isolated). For the starless/protostellar distinction, we treated uncertain classifications as distinct categorical variables. The separate classification of uncertain objects essentially eliminated these objects from our analysis, but allowed us to check for any unusual properties these uncertain objects might have had.

As with many traditional statistical tests, ANOVA assumes that the underlying variables are normally distributed, but this is generally not the case with our measured variables. However, in many cases the variables can be transformed with a logarithmic transformation to be approximately normal, and this allows us to interpret the statistical significance of our findings within the ANOVA framework. Differences between means of logarithmically transformed variables are expressed as ratios of the median values because the transformation is monotonic (preserves the median) and the transformed distribution is roughly normal so the median approximates the mean. The differences between the category variables are generally small compared to the spread within individual populations. 

To derive realistic confidence intervals we used Tukey's Honest Significant Differences (HSD) \citep[][pp. 97-98]{Yandell:1997}, which account for the fact that we were approaching the data set without specific comparisons to test, but were rather testing the significance of all possible pairings. This is the most conservative approach to setting confidence intervals. 

Our ANOVA procedure makes the assumption that the variance within groups are the same. We tested this assumption using Levene's test \citep{Levene:1960}, and cases where significant non-uniformity where encountered are mentioned in the text and summary table. These significance levels and confidence intervals should be treated with greater suspicion, but the ANOVA tests are relatively insensitive to non-uniform variance \citep[see][]{Levene:1960}, so the general sense of the result is probably unaffected. 

We performed these tasks in the R statistical package. The results are summarized in Table \ref{ANOVAtable} and will be discussed in more detail for each variable in the following sections.

\begin{deluxetable*}{ccccp{1.9in}p{0.9in}c}
\tablewidth{0pc}
\tabletypesize{\scriptsize}
\tablecaption{Summary of ANOVA Results}
\tablehead{Variable & $f(x)\tablenotemark{a}$ & Cluster Significant? & Protostar Significant? & Statement & Comments & \# of Cores} 
\startdata
T$_\mathrm{kin}$ & log & Yes, P = 6\e{-5} & 0.01 & Med(T$_K$) Starred:Starless = 1.11 & Long-tailed after& 128 \\
                  &        &                &        & 95\% CI(1.01,1.23)                              &  log transform & \\
                                    &        &                   &        & Med(X-CCS) Cluster:Isolated = 1.16                           &   &  \\ 
                  &        &                   &        & 95\% CI(1.08,1.24)                              & &  \\ 
                  \\
T$_\mathrm{ex}$ &         &  No            & Yes, P = 1\e{-7}& T$_\mathrm{ex}$ Starred - Starless= 1.26 K & Normal, unequal & 118 \\
                  &        &                   &         & 95\% CI(0.73,1.79)                              & variance\tablenotemark{b}& \\
                  \\
N(H$_2$) &  log &  No            & Yes, P = 3\e{-10} & Med(N(H$_2$)) Starred:Starless = 2.82 & Normal after & 181 \\
                  &        &                   &       & 95\% CI(1.92,4.16)                              & log transform& \\ 
                  \\
X-NH$_3$     &  log  &  Yes, P =  0.02          & No & Med(X-NH$_3$) Cluster:Isolated = 0.56 & Skewed after & 178 \\
                  &        &                    &        & 95\% CI(0.35,0.91)                              & log transform& \\ 
                                    \\

X-CCS     &  log  &  Yes, P = 8e{-8}            &  Yes, P = 2\e{-5}& Med(X-CCS) Starred:Starless = 0.37 & Uses 1$\sigma$ limits & 181 \\
                  &        &          &       & 95\% CI(0.22,0.62)                              & as data points.& \\ 
                  &        &                   &        & Med(X-CCS) Cluster:Isolated = 0.32                             & Similar results  &  \\ 
                  &        &                   &        & 95\% CI(0.21,0.48)                              & without limits& \\ 
                  \\
V$_\mathrm{NT}$ (NH$_3$) &  &  No            & Yes, P = 4\e{-3}& V$_\mathrm{non-thermal}$ Starred - Starless = 0.025 km/s & Normal & 83 \\
                  &        &                   &       & 95\% CI(0.003,0.048)                              & & \\ 
                  \\
V$_\mathrm{NT}$ (CCS) &  &  No            & No & No significant influence & Still long-tailed, no CCS in protostellar + cluster & 59 \\
                  &        &                   &        &                               & & \\ 
                 \\
Mass & log & No &      Yes, P = 0.05   & Med(Mass) Starred:Starless = 1.32 & Normal after & 123 \\
                    &       &       &         & 95\% CI(1.00,1.76)                             & log, unequal & \\
                    &	 &	  &	    & 							    & variance\tablenotemark{c}& \\
                    \\
$\alpha$ &  log &  No            & No & No significant influence & Normal after log & 69 \\
                  &        &                   &        &                           & transform& \\ 

                  \tablenotetext{a} {The transformation applied to make the distribution more normal. All logs are log base 10.}
		\tablenotetext{b} {Using Levene's test we find the excitation temperatures of protostellar and starless cores to have unequal variance at the P = 0.02 level.}
		\tablenotetext{c} {Using  Levene's test we find the log(Masses) of protostellar and starless cores to have unequal variance at the P = 0.01 level.}

\label{ANOVAtable}
\enddata
\end{deluxetable*}

\subsection{Sensitivity to Random and Systematic Temperature Errors}

Most of the derived variables we study herein rely in some way on using a temperature (T$_\mathrm{kin}$) derived from modeling the NH$_3$ spectra to convert observed quantities into physical units. This dependence on temperature can be divided into two cases:

1) The distributions which use only the NH$_3$ and CCS data (T$_\mathrm{ex}$, T$_\mathrm{kin}$, and both non-thermal linewidths) require that our determination of temperature accurately measures the temperature of the bulk (H$_2$) gas. The most serious error possible is a systematic one. 

2) The distributions which also rely on the Bolocam maps (N(H$_2$), M$_\mathrm{dust}$, $\alpha$, X(CCS) and X(NH$_3$)) require that T$_\mathrm{kin}$ from NH$_3$ adequately approximates T$_\mathrm{dust}$. This is an estimate and may be biased in some way. We show in Appendix B that the best estimate for this discrepancy has a small influence on our results. Nonetheless, for this half of the analysis the probabilities quoted are valid only under the assumption that T$_\mathrm{kin}$ for NH$_3$ adequately approximates T$_\mathrm{dust}$.

As the two simplest quantities to assess, we perform experiments on the non-thermal linewidth of NH$_3$ (case 1) and the total column density (case 2) in Appendix B. These tests show that our results are very robust against the influence of random errors on the temperature. This is due to the high-quality of our spectra, which allow a very precise estimate of the temperature. The non-thermal linewidth result is robust against systematic errors in the temperature determination. Systematic errors in the temperature determination can be constructed which eliminate the distinctions we see in quantities which rely on the dust emission map, particularly if these systematic differences work in different directions for the two populations (i.e. protostellar/starless). However, under our best estimate of the possible systematic error our results are affected by only a small amount. 

We use these experiments to infer the influence of random and systematic temperature errors on our other variables.

The non-thermal linewidths of CCS will behave much like the non-thermal linewidths of NH$_3$; they will be influenced only slightly by either sort of temperature error.

Errors in total column density will propagate into the fractional abundances of NH$_3$ and CCS mostly in the obvious way (increasing the total column density of material decreases the fractional abundance). A bad fit to the NH$_3$ could trade off some NH$_3$ column for some T$_\mathrm{kin}$, but as our spectra are generally of high quality and our fits are generally quite good,  we do not think this is likely to be a problem. Masses (which come directly from the dust map) will obviously be affected in the same was a column density. The virial parameter ($\alpha =  M_\mathrm{dyn}/M_\mathrm{dust}$) has a double dependence on T$_\mathrm{kin}$. However, since our experiments in Appendix B show non-thermal linewidths to be influenced by only a small amount by reasonable errors on T$_\mathrm{kin}$, the dominant effect will be in M$_\mathrm{dust}$. Thus if systematic errors are present which cause N(H$_2$) to be overestimated for a given object, $\alpha$ will underestimated, and vice-versa.

\section{Results}
\label{Results}

\subsection{Temperature}
\label{Temp_star}

\subsubsection{Kinetic Temperatures: T$_\mathrm{kin}$}

\begin{figure}
\includegraphics[scale=0.8]{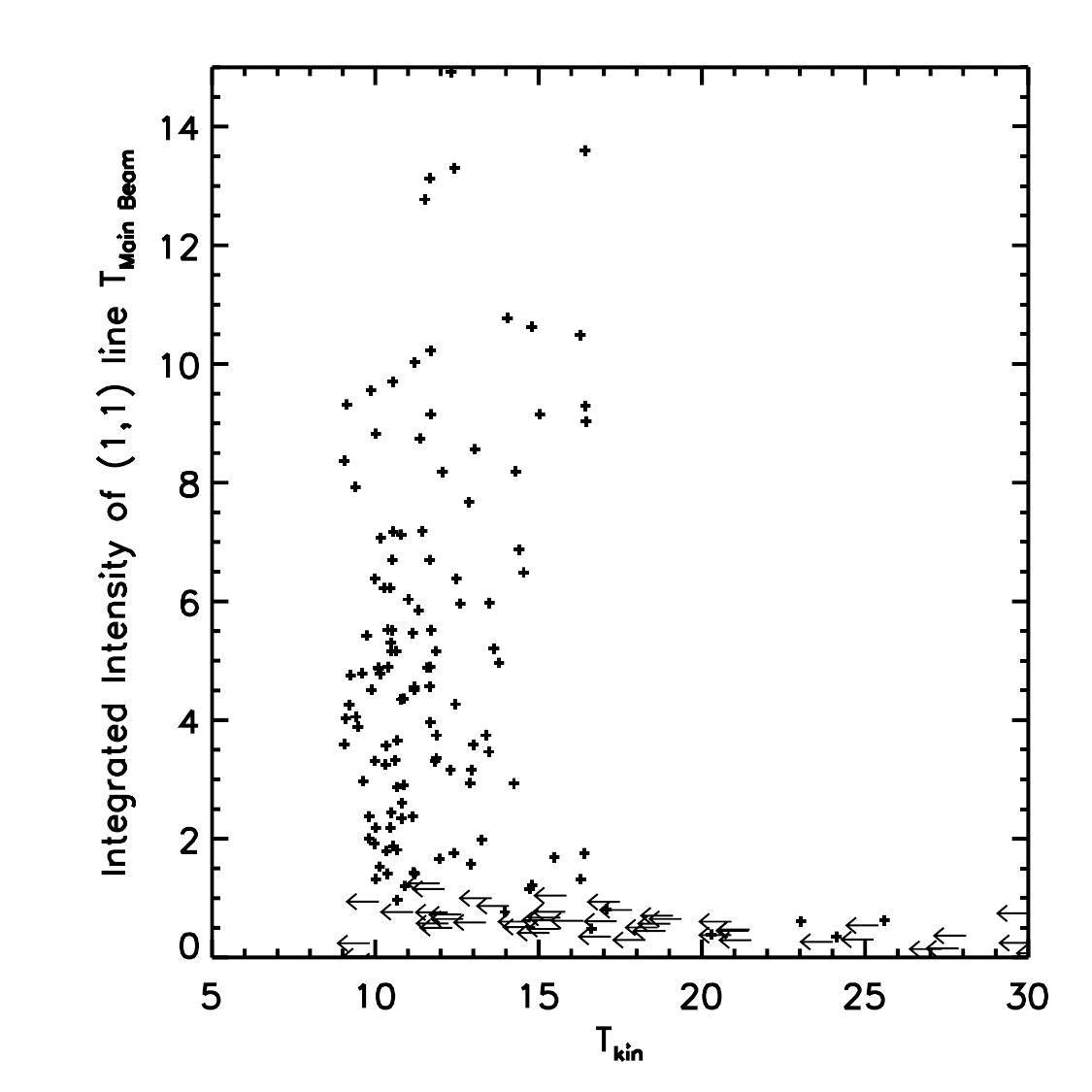}
\caption{ T$_\mathrm{kin}$ versus the integrated intensity of the (1,1) line. Good solutions for T$_\mathrm{kin}$ are shown as plus symbols and 3 $\sigma$ upper limits on T$_\mathrm{kin}$ \citep[derived in][using the noise in each individual spectrum]{Rosolowsky:2008} are shown for the rest of the objects.}
\label{TempLimits}
\end{figure}

The kinetic temperature of cores were fit from the (1,1) and (2,2) transitions of NH$_3$ and depend on the relative strengths of these two transitions. They are thus determined only for objects with a detected (2,2) line. Our ability to detect this line depends on the NH$_3$ column and temperature of the core. Thus, we are generally insensitive to very low kinetic temperatures, and at low temperatures we are biased against being able to solve for the temperature of diffuse or low-density material. This may not be a significant problem for our survey, as we illustrate in Figure \ref{TempLimits}, which shows T$_\mathrm{kin}$ versus the integrated intensity of the (1,1) line. Good solutions for T$_\mathrm{kin}$ are shown as plus symbols and 3 $\sigma$ upper limits on T$_\mathrm{kin}$ \citep[derived in][using the noise in each individual spectrum]{Rosolowsky:2008} are shown for the rest of the objects. These upper limits are consistent with our survey not being significantly biased against detecting cores with T$_\mathrm{kin}$ below 9 K. That is, the (1,1) is always quite weak for objects without a (2,2) detection. We cannot definitively prove that there is not a population of low temperature cores with weak (1,1) lines as well, but we do not consider this likely, partly for the theoretical reason that low density cores would not self-shield against radiation and thus would be warmer. 

This does not imply that the central temperatures of all these cores are above 9 K. Our model fits a single temperature to the NH$_3$ spectra. This is a good, but imperfect model. We will overestimate kinetic temperatures for cores with a significant temperature gradient due to temperature variation along the line of sight as well as structure on the sky which is smaller than our beam. The contribution from the warm material will tend to dominate over the cold. Thus, our results are not inconsistent with cores being colder than 9 K in the center, as found in TMC-1c by \citet{Schnee:2005} where the central temperature was 6-7 K in the central 0.03 pc, a radius somewhat smaller than our beam.

\begin{figure*}
\includegraphics*{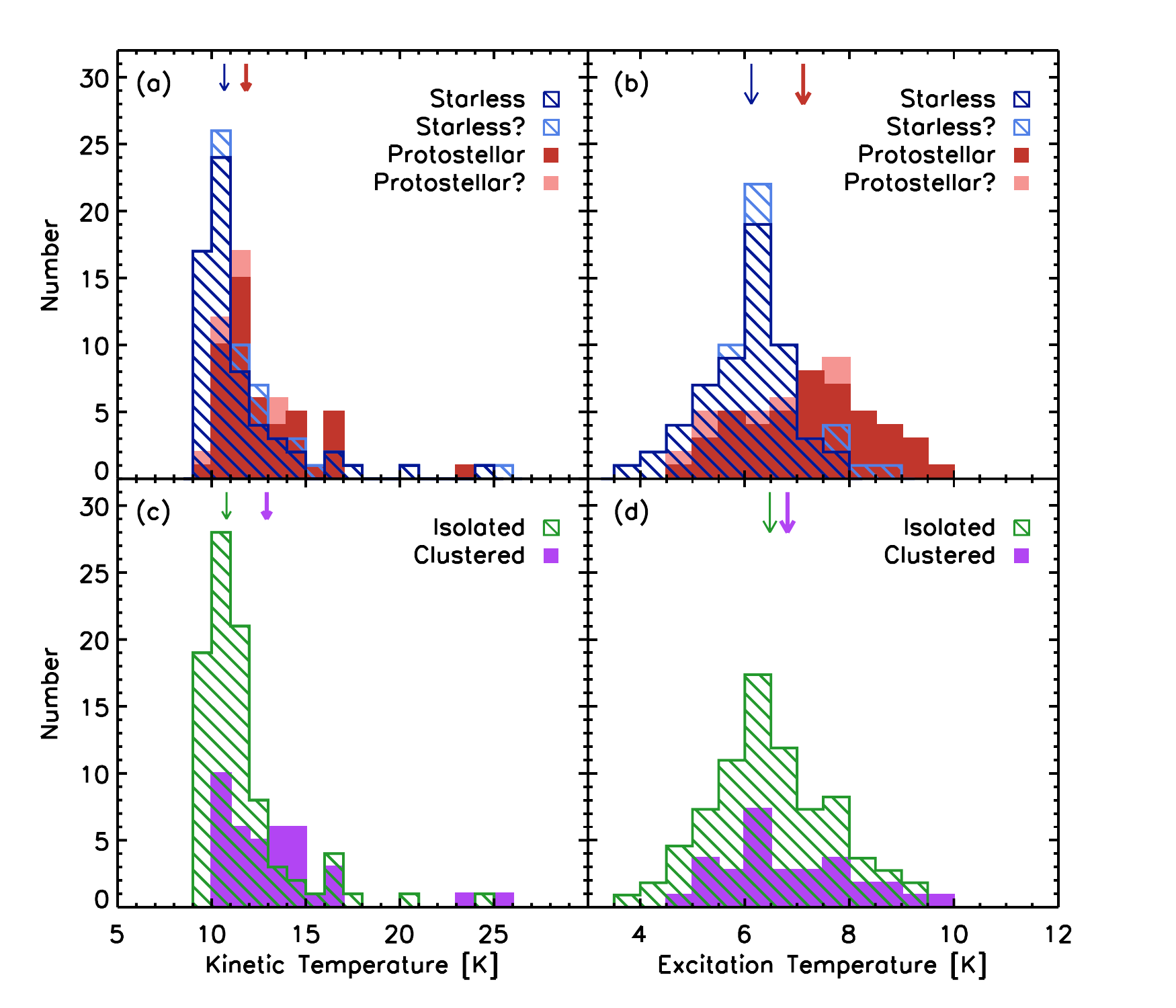}
\caption{Histograms of kinetic and excitation temperature for ammonia cores broken down by protostellar content and location within a cluster. Panels a and b show starless (blue/dashed) and protostellar (red/solid) cores. Uncertain identifications are identified by lighter colors and are stacked on top of the certain identifications. Panels c and d show isolated (green/dashed) cores and cores inside of clusters (purple/solid). Arrows show median values of the distributions with the thicker arrow pointing to protostellar and cluster means.}
\label{Temp}
\end{figure*}

Figures \ref{Temp}a and \ref{Temp}b show the temperature distributions of our cores broken down into categories and are typical of the histograms displayed throughout this paper. Here the dashed-blue bars are starless (think empty/cold) while the solid-red bars are cores associated with a protostar (think full/hot). Certain identifications are displayed in heavy or darker lines, while the lighter blue and pink portions correspond to data points with uncertain assignment. The two main categories (in this case starless/protostellar) are overlain on each other, while the uncertain points are stacked on top of the bars rather than being displayed as additional layers. Objects are included in the plots if it is possible to derive a kinetic temperature from the ammonia spectra, which requires a detection of both the (1,1) and the (2,2) transition. If the (2,2) line is not detected, \citet{Rosolowsky:2008} provides an upper limit on the temperature. These are not included on this plot, but see Figure~\ref{TempLimits}.

Virtually every point below 10 K is starless, but the majority of protostellar cores are also relatively cool, between 10 and 15 K. They are thus slightly cooler than the 15 K assumed in \citet{Enoch:2008}. Surprisingly, 3 of the 4 hottest cores (T$_\mathrm{kin}$ $>$ 20 K) are starless, so not all starless cores are cold. These objects are, however, at low column density, with relatively weak detections of both the (1,1) and (2,2) line (see Figure~\ref{TempLimits}). \citet{Schnee:2008} provide a detailed comparison of these same NH$_3$ temperatures with dust temperatures for starless cores in Perseus and study the correlation of these temperatures with other variables, but are unable to isolate any variable which strongly influences the temperature of starless cores other than that isolated cores are colder than ones in clusters.

The distribution of kinetic temperatures is non-Gaussian, though this may be in part because the SCUBA/Bolocam surveys which selected most of our cores are flux limited, which corresponds to a combined temperature/mass limitation. In addition, as noted above, we are biased against being able to solve for the temperature for very cold objects due to non-detection of the (2,2) transition. The isolated cores appear to form an approximately normal distribution centered at 10 K, while the cores in clusters have a roughly uniform distribution between 10 and 15 K. Of the high-temperature outliers mentioned above, only two are in clusters. There remain two isolated starless cores with temperatures above 20 K.

Kinetic temperature is also a good example of why the ANOVA formalism is useful for distinguishing between the influence of two categorical variables. The general trend is towards colder temperatures in both starless cores (Figure \ref{Temp}a) and isolated cores (Figure \ref{Temp}b), and we might reasonably expect both the presence of a protostar and the surrounding cluster environment to warm a core. This variable also illustrates how overlapping these populations are: variations within populations are larger than the differences between them. 

For the ANOVA test we applied a logarithmic transformation, which made the distribution closer to normal. However, the resulting variable was still long-tailed as visualized on a Normal Q-Q plot (not shown). We found that cluster environment is highly significant (P = 6\e{-5}), while stellar content is somewhat less significant (P=0.01). Clustered cores are 1.16 times warmer than isolated ones (95\% Confidence Interval:[1.08,1.24], while protostellar cores are 1.11 times warmer than starless cores (95\% Confidence Interval:[1.01,1.23]). The combination of cluster environment and a protostar does not increase the temperature beyond the additive increase from these two individual factors.

\subsubsection{Excitation Temperatures: T$_\mathrm{ex}$}

Excitation temperatures (T$_\mathrm{ex}$) are only reliably fit at moderate optical depths \citep[see][]{Rosolowsky:2008}. For this sample, the fit temperatures were normally distributed with no significant outliers. Stellar content is highly significant, while cluster environment is not. Excitation temperatures are 1.26 K higher in protostellar cores (95 \% Confidence Interval:[0.73,1.79] P = 6\e{-7}). Figure~\ref{Temp}b shows that the protostellar cores have larger variance than the starless cores, a fact confirmed by Levene's test at the P = 0.02 level. Nonetheless, we remain confident that the protostellar cores do have higher excitation temperatures and propose that this difference arises because protostellar cores have higher volume densities.

\begin{figure}
\includegraphics[scale=0.8]{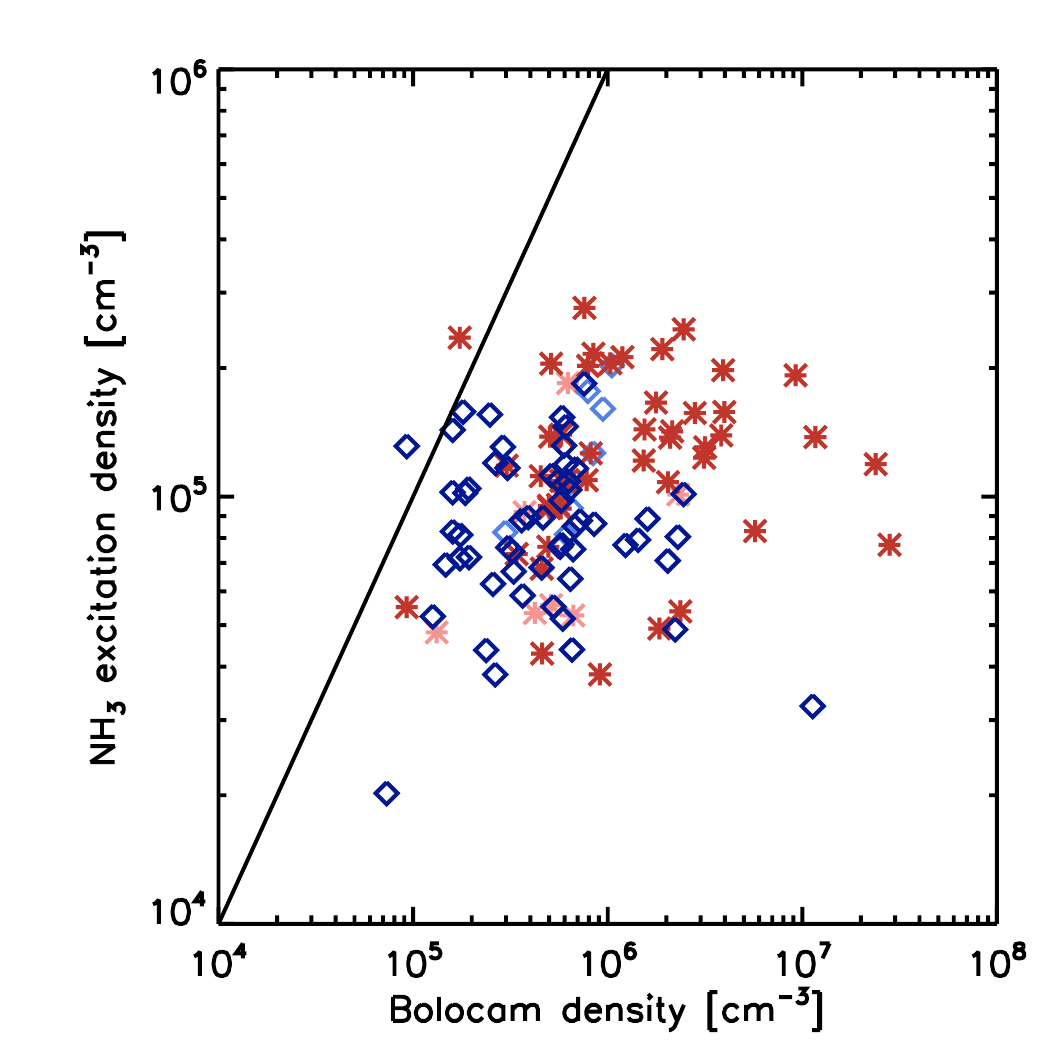}
\caption{Comparison of volume density estimates for protostellar (red/filled) and starless (blue/open) cores by two independent methods. The ordinate shows an estimate from comparing T$_\mathrm{kin}$ to T$_\mathrm{ex}$ from the NH$_3$ data. The abscissa shows an estimate from the masses and radii derived from the Bolocam survey of \citet{Enoch:2006}. The line shows equality. Despite the disagreement, both methods agree that protostellar cores are denser than starless cores.}
\label{VolumeDensity}
\end{figure}

\subsection{Volume Densities (n)}

We can make an estimate of the volume density for each core under the assumption of LTE from T$_\mathrm{ex}$, T$_\mathrm{kin}$, and $\tau$ as \citep{Swift:2005,Caselli:2002}
\begin{equation}
n = \frac{(J(T_\mathrm{ex}) - J(T_\mathrm{cmb}))k J(T_\mathrm{kin})}{h \nu_{1,1} (1 - J(T_\mathrm{ex}))} \times n_\mathrm{crit} \times \beta,
\end{equation}
where
\begin{equation}
J(T)= \frac{h \nu_{1,1} }{k (1 - e^{-\frac{h  \nu_{1,1}}{k T}})},
\end{equation}
and T$_\mathrm{cmb}$ = 2.73 K. The escape probability, $\beta$, is roughly estimated for as slab as $\beta = (1 - e^{-\tau})/\tau$ \citep{Swift:2005}, an estimate which works out to roughly 0.5 for our typical values of $\tau$. We take $\tau$ as the maximum $\tau$ (estimated as $\tau_{(1,1)} \times 0.233$ since 0.233 is the maximum opacity in any of the hyperfine components). We use n$_\mathrm{crit}$ = 2\e{4}, a fairly standard value \citep{Myers:1999}.

A second estimate of the volume density comes from simply dividing the mass of a core by the volume, as estimated from the radius. This is the approach taken in \citet{Enoch:2008}, though we modify those masses based on our estimates of T$_\mathrm{kin}$ (see \S~\ref{MassStability}). We compare these densities to those estimated from NH$_3$ in Figure~\ref{VolumeDensity}. These two estimates systematically disagree, in the sense that the Bolocam density estimates are roughly a factor of ten larger, and show very little correlation with each other. There are plausible explanations, however. First, we only see NH$_3$ emission from certain regions of a core, so it is not surprising that we see densities close to the characteristic density \citep{Swade:1989}. Second, the Bolocam masses may be systematically wrong due to uncertainty in the value of the dust emissivity at 1.1 mm, $\kappa$. Third, the critical density is calculated as
\begin{equation}
n_\mathrm{crit} = \frac{A_{ul}}{C_{ul}}
\end{equation}
where the Einstein A$_{ul}$ coefficient for the (1,1) transition is well established as 1.67\e{-7} s$^{-1}$ \citep{Ho:1983}, but C$_{ul}$ (the collisional cross section) is only poorly known. Calculations of its value range from 8.6\e{-11} cm$^3$ s$^{-1}$ in \citet{Danby:1988} (for H$_2$) to 5\e{-12} cm$^3$ s$^{-1}$ in \citet{Machin:2005} (for He, both values are for 15 K gas). Fourth, the Bolocam densities represent core-average densities, while the NH$_3$ density calculation is for the portion of the core within the beam. As most Bolocam cores are larger than our GBT beam and we typically point at the center (and presumably densest portion) of the core, this will tend to decrease the Bolocam densities on average. Fifth, Bolocam observations partially filter out extended dust emission from core envelopes which is still detected in NH$_3$. Taken together, these five factors can easily explain the systematic offset and lack of correlation in these two density estimates.

Regardless of which density determination is more accurate, protostellar cores appear to have higher volume density than starless cores. Nonetheless, because of the imprecision of this estimate, we do not attempt further statistical tests on this variable.

\subsection{Total Column Density: N(H$_2$)}
\label{ColDensity}

We calculate the column density of dust and use this to estimate total column density. This is done by measuring the flux at the appropriate pixel in the Bolocam map and converting to N(H$_2$) using

\begin{equation}
N(H_2) = \frac{S_{\nu}^\mathrm{beam}}{\Omega_{A} \mu_{H_2} m_H \kappa_{\nu} B_{\nu} (T)},
\end{equation}
where $S_{\nu}^\mathrm{beam}$ is the flux per beam, $\Omega_{A}$ is the beam solid angle, $\mu_{\mathrm{H_2}}$ is the mass per hydrogen molecule, $\kappa_{\nu}$ is the dust opacity, and $B_{\nu} (T)$ is the Planck function. We used $\mu_{\mathrm{H_2}} = 2.8$, $\lambda = 1120\, \mu$m, $\theta_\mathrm{HPBW} =$ 31\arcsec,  and $\kappa_{\nu} = 0.0114 $ [cm$^{2}$ g$^{-1} $] \citep{Ossenkopf:1994}. This value of $\kappa_{\nu}$ corresponds to the wavelength-interpolated \citet{Ossenkopf:1994} value for dust grains with thin ice mantels in dense regions (n = 10$^{6}$ cm$^{-3}$). This opacity differs from that adopted in \citet{Andre:1996}, which uses an opacity a factor of 2 lower, consistent with the recommendation of \citet{Henning:1995} for lower ($<$ 10$^{5}$ cm$^{-3}$) density regions. As discussed above, the densities of these cores is uncertain, and certainly changes throughout the core, making the opacity adopted a significant source of uncertainty. With these values, column densities scale like
\begin{equation}
\label{T_N_relation}
N \sim e^{12.85/T} - 1.
\end{equation}

The Bolocam map covers virtually all our ammonia pointings. Although 10 cores fall outside the Bolocam survey, 7 of these are non-detections. For consistency, points without a Bolocam flux are not included, even if an estimate of the column density is available from another source. 

This calculation requires a temperature, for which we assume the NH$_3$ kinetic temperature. In general, we expect strong coupling between dust and gas temperature at such high densities \citep{Goldsmith:1978}. In the case of protostellar cores, there may be strong temperature and chemistry gradients weakening the connection between ammonia and dust temperatures. Errors in T from this gas-dust coupling assumption will propagate according to Eqn. \ref{T_N_relation}.

For the small number (37) of objects where we had a NH$_3$ (1,1) detection but were not able to determine a temperature from the ammonia (due to an undetected (2,2) line), we assume T = 11 K, which is roughly the mode and the median of the values in Figure \ref{Temp}. 

\begin{figure}
\includegraphics*{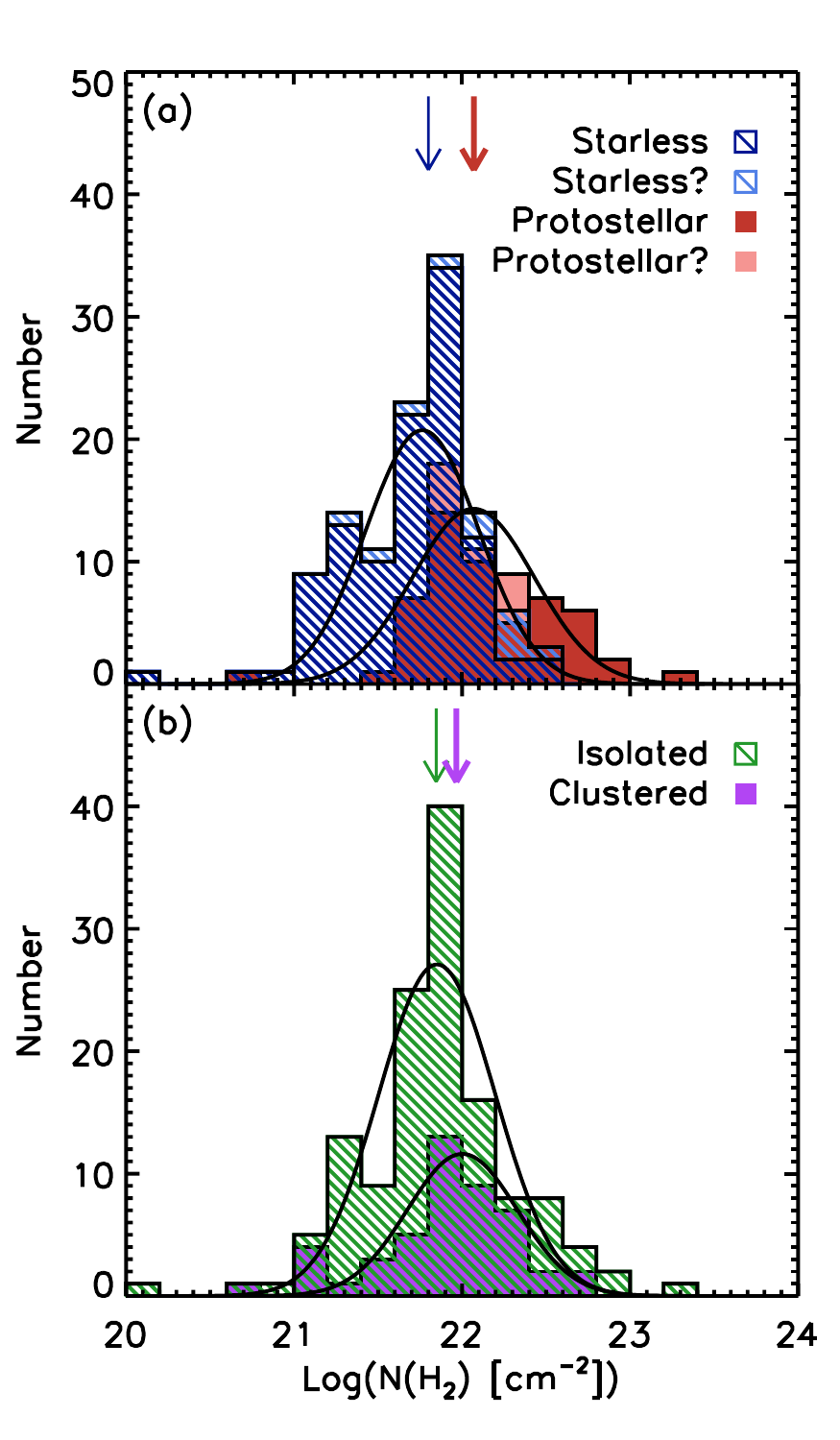}
\caption{Histograms of column density at ammonia pointings as derived from the Bolocam flux at the position of the GBT pointing and temperatures derived from NH$_3$. Panel a shows starless (blue/dashed) and protostellar (red/solid) objects. Uncertain identifications are identified by lighter colors and are stacked on top of the certain identifications. Panel b shows isolated (green/dashed) cores and cores inside of clusters (purple/solid). Arrows show median values of the distributions. Gaussian fits are shown and tabulated in Table~\ref{Gauss}}
\label{NH2}
\end{figure}

The distribution of measured column density is non-Gaussian, but becomes essentially normal under the logarithm transformation, reminiscent of the log-normal distribution of column densities throughout the cloud as a whole \citep{Ridge:2006,Goodman:2008}, but significantly denser. These values generally represent the distribution of the highest density peaks in the cloud, as these are the regions bright enough to be seen in the Bolocam map used to select the majority of our targets. We fit Gaussians to these distributions in Figure~\ref{NH2} and tabulate the fit parameters in Table~\ref{Gauss}.
We use the conversion 
\begin{equation}
N(H_2) = 9.4\e{20} \mathrm{cm}^{-2} \mathrm{(A_V/mag)}
\end{equation}
from \citet{Bohlin:1978} to convert our N(H$_2$) values into an A$_V$ for comparison.

In the study \citet{Goodman:2008}, a mean value of A$_V$ = 1.18 was found for an extinction map of the whole cloud.  The mean A$_V$ for our entire sample is 8.4 for this fit, confirming that this distribution represents the high-density portions of the cloud.

In our ANOVA formalism we see similar results. We derive lower column densities for starless cores than for protostellar cores (Figure \ref{NH2}), but find no significant difference in clustered versus isolated environs. The median protostellar core has 2.82 times higher column density than the median starless core (95\% Confidence Interval:[1.92,4.16] P = 3\e{-10}). Cluster environment has a statistically insignificant influence on dust column density.

\begin{deluxetable}{lcc}
\tablewidth{0pc}
\tablecaption{Gaussian Fits to Logarithmic Column Density Distributions}
\tablehead{ \colhead{Category} & \colhead{Mean Log(N(H$_2$))} & \colhead{$\sigma$ Log(N(H$_2$))} }
\startdata
      All  &    21.90   &  0.32 \\
      Starless  &    21.76  &   0.34 \\
      Protostellar    &  22.07  &   0.36  \\
      Isolated   &   21.85  &   0.35  \\
      Cluster  &    22.00  &   0.33   \\
\enddata
\label{Gauss}
\end{deluxetable}

\subsection{Chemistry}
\label{Chemistry}

The derivation of NH$_3$ and CCS column densities are explained in \citet{Rosolowsky:2008}, and obviously require detections of the line in question. As the Bolocam beam is nearly the same size as the GBT beam (31\arcsec), we derive a fractional abundance by simply dividing the species' column density by N(H$_2$). The dominant uncertainty in these chemical calculations is a possible mismatch between T$_\mathrm{kin}$ and T$_\mathrm{dust}$ which would produce errors in our estimate of N(H$_2$) of the form described in \S\ref{ColDensity}.

\subsubsection{Fractional Abundance of NH$_3$: X-NH$_3$}

We detect NH$_3$ toward 178 of our 193 pointings, so the influence of missing the cores with the lowest abundances of NH$_3$ (a form of censorship bias) is probably relatively small. While the fractional abundance of ammonia does not appear to be a strong function of stellar content, clustered cores generally are less abundant in NH$_3$  (Figure~\ref{AbundanceHist}a \& c). Presumably it is the combined radiation environment in the cluster that is destroying NH$_3$ rather than the influence of an individual embedded protostar. The ANOVA test verifies this distinction. Under a logarithmic transformation, the initially skewed fractional abundance of NH$_3$ remained somewhat skewed. Only cluster environment proved to be significant, in the sense that the median clustered core had 56\% the fractional NH$_3$ abundance of the median isolated core (95\% Confidence Interval:[0.35,0.90], P = 0.02).

\subsubsection{Fractional Abundance of CCS: X-CCS}
\label{FracAbundanceOfCCS}

\begin{figure*}
\includegraphics*{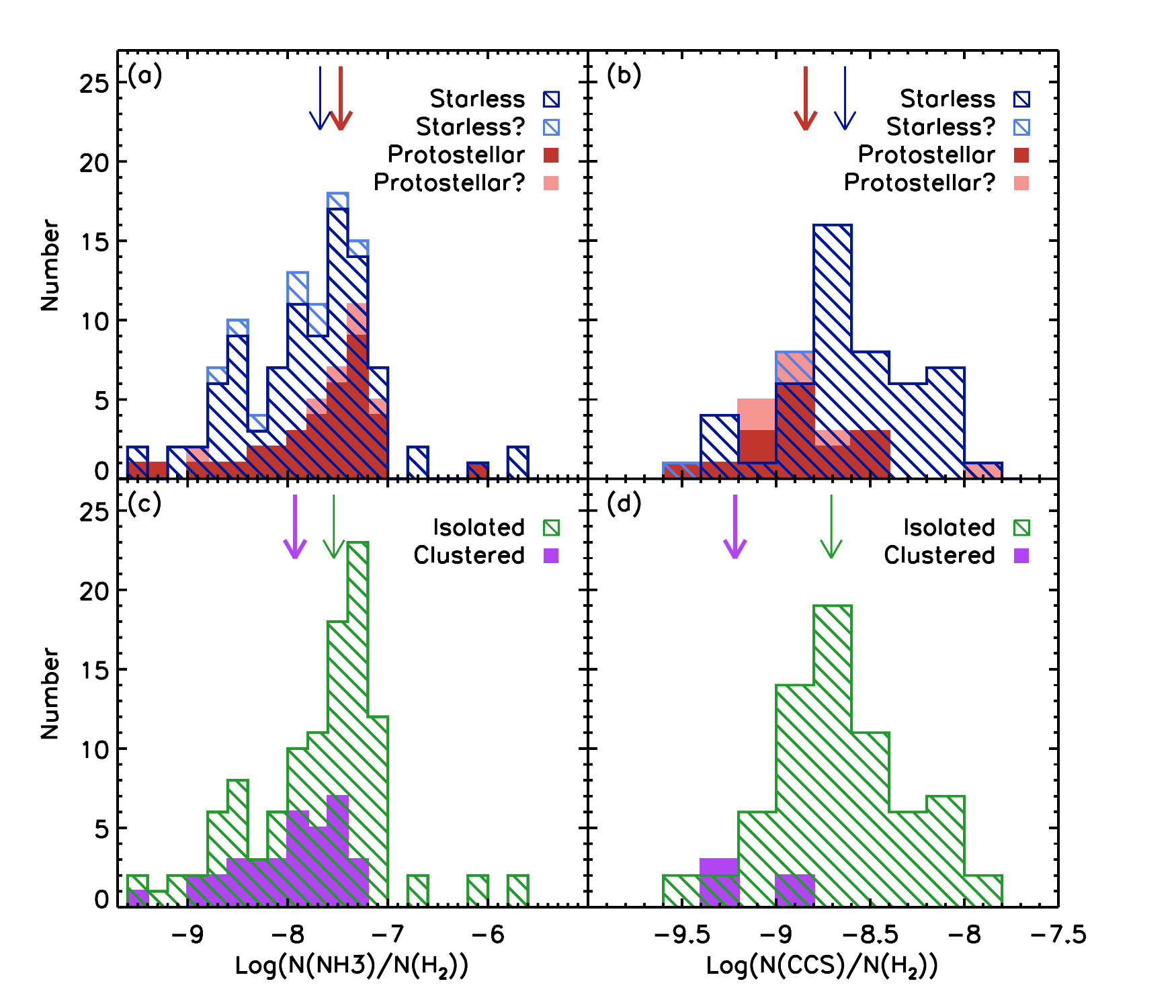}
\caption{Histograms of fractional abundances of CCS and NH$_3$ broken down by protostellar content and location within a cluster. Panels a and b show starless (blue/dashed) and protostellar (red/solid) cores. Uncertain identifications are identified by lighter colors and are stacked on top of the certain identifications. Panels c and d show isolated (green/dashed) cores and cores inside of clusters (purple/solid). Arrows show median values of the distributions.}
\label{AbundanceHist}
\end{figure*}

In the case of CCS, censorship bias becomes a significant problem. Less than half our objects have a reliable CCS detection (94 out of 193), and there are no protostellar cores in clusters with CCS detections (Figure~\ref{AbundanceHist}b \& d). In fact, only five cores within a cluster have a CCS detection, and they are all quite weak. In order to present a comparable analysis, we assigned 1$\sigma$ upper limits to the fractional abundance of CCS based on the noise in each individual spectrum, the temperature, and an estimate of the width of the line from the linewidth for NH$_3$ at that pointing. Then, treating these upper limits as data points alongside our genuine detections, we ran the ANOVA test. The fractional abundance of CCS became roughly normal after a logarithmic transformation. In this case, both category variables were highly significant (still without a significant interaction effect). We estimate that the median protostellar cores had 37\% the fractional CCS abundance of the median starless core (95\% Confidence Interval:[0.22,0.62] P = 2\e{-5}), while the median clustered core had 32\% the fractional CCS abundance of the median isolated core (95\% Confidence Interval:[0.21,0.48] P=8\e{-8}). This method overestimates the CCS abundance in clusters and protostellar cores, since more of these objects have 1 $\sigma$ upper limits (which will typically be higher than the true values). Thus, the fractional abundance of CCS in clusters and protostellar cores is probably less than presented here, and even more distinct from the fractional abundances of CCS in isolated and starless cores.

This distribution is one case in which the division of cores into ``influenced'' and ``not-influenced'' by our association with 24 micron sources (see Appendix A) produces an interestingly different result. The contrast in fractional abundance of CCS is more pronounced, with virtually all ``non-influenced'' cores lying at high abundance. This could be due to radiation from YSO influencing the chemistry in the outer layers of cores, even if that source is not actually embedded within the core.

\begin{figure}
\includegraphics*[scale=0.7]{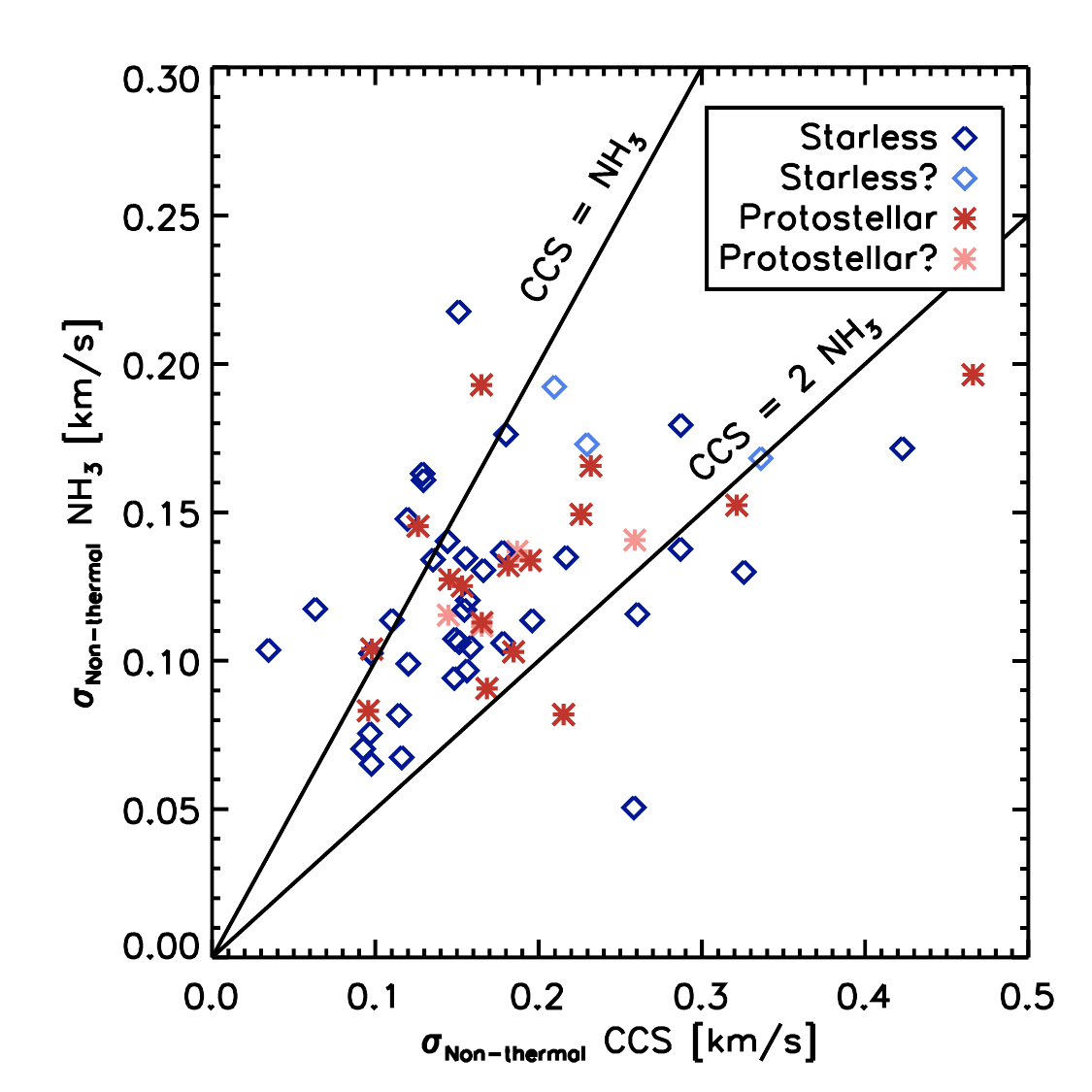}
\caption{Non-thermal linewidths in CCS versus NH$_3$ for starless and protostellar cores. Non-thermal motions are larger in CCS than NH$_3$, indicating that these molecules are tracing different physical regions within the core. Histograms of this plot projected onto the axes are presented in Figures \ref{NonThermalWidths}a and b.}
\label{NonThermalVsNonThermal}
\end{figure}

\begin{figure}
\includegraphics*[scale=0.7]{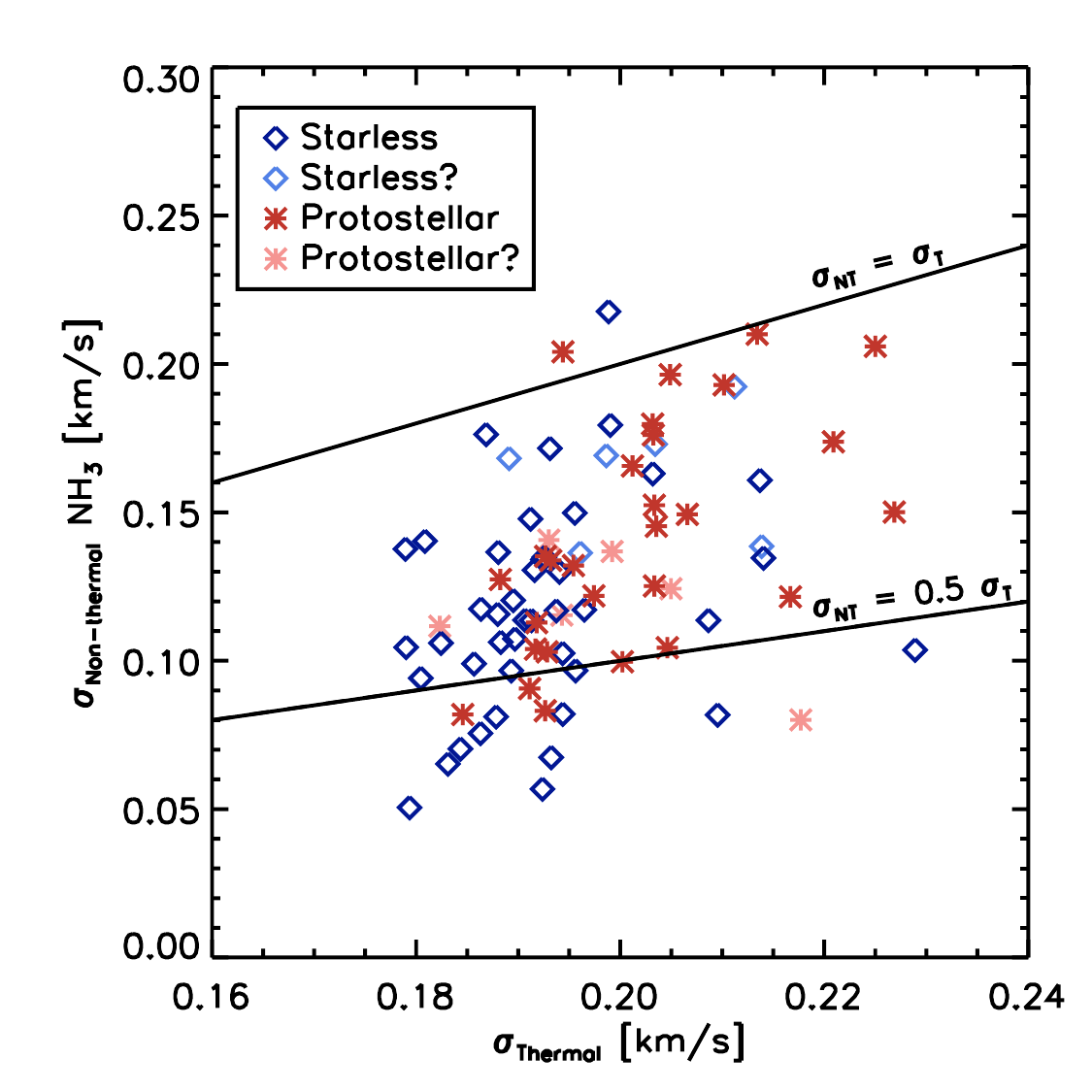}
\caption{Thermal linewidths for H$_2$ versus non-thermal linewidths leftover in the NH$_3$ profile for starless and protostellar cores. Lines demarcate cores with non-thermal linewidths equal to thermal and cores with non-thermal linewidths half of thermal. Most cores are quiescent; their non-thermal linewidths are less than thermal. A histogram of this plot projected onto the ordinate is presented in Figure \ref{NonThermalWidths}a.}
\label{ThermalVsNon}
\end{figure}

\begin{figure*}
\includegraphics*{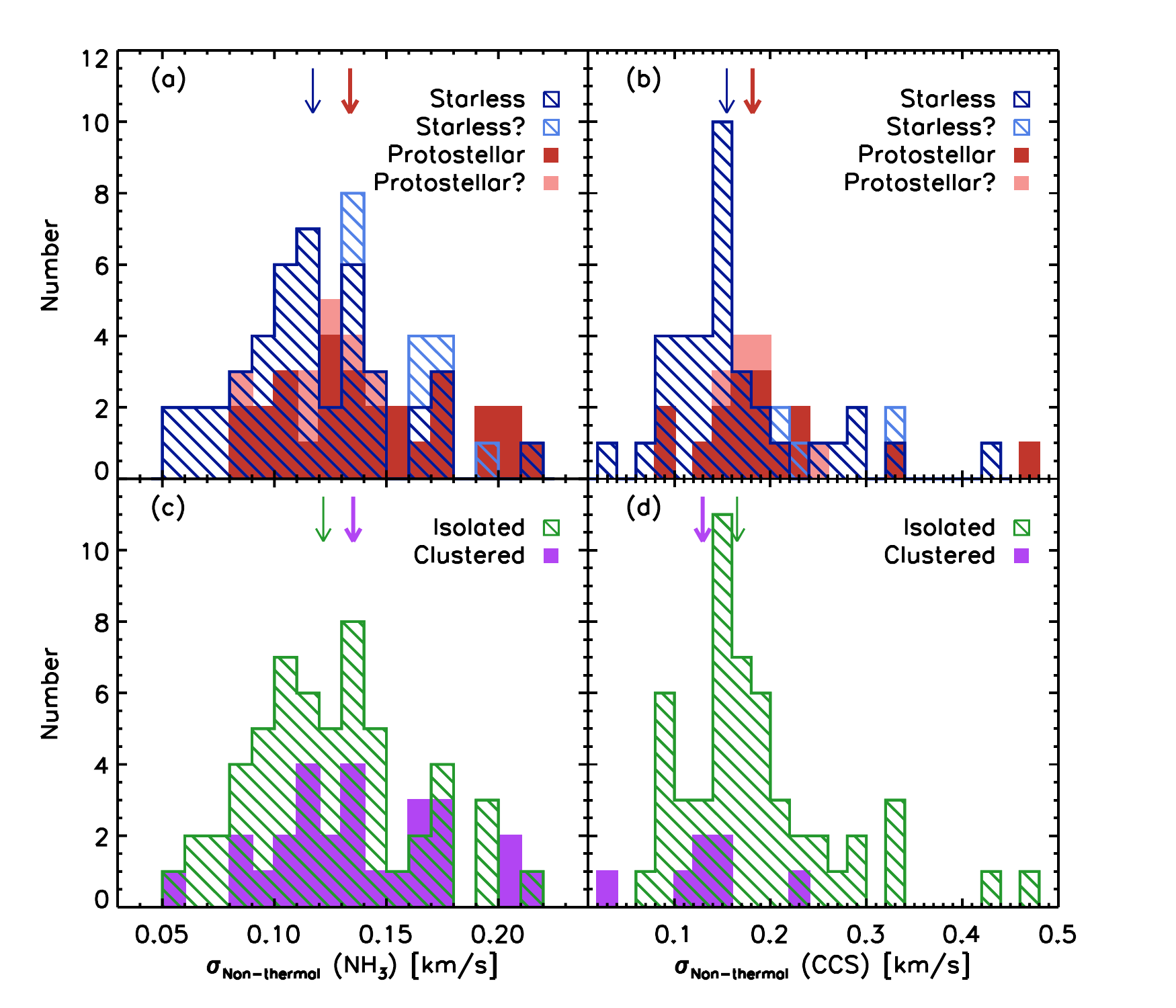}
\caption{Histograms of non-thermal linewidths for CCS and NH$_3$ broken down by protostellar content and location within a cluster. Panels a and b show starless (blue/dashed) and protostellar (red/solid) cores. Uncertain identifications are identified by lighter colors and are stacked on top of the certain identifications. Panels c and d show isolated (green/dashed) cores and cores inside of clusters (purple/solid). Arrows show median values of the distributions.}
\label{NonThermalWidths}
\end{figure*}

\begin{figure}
\includegraphics{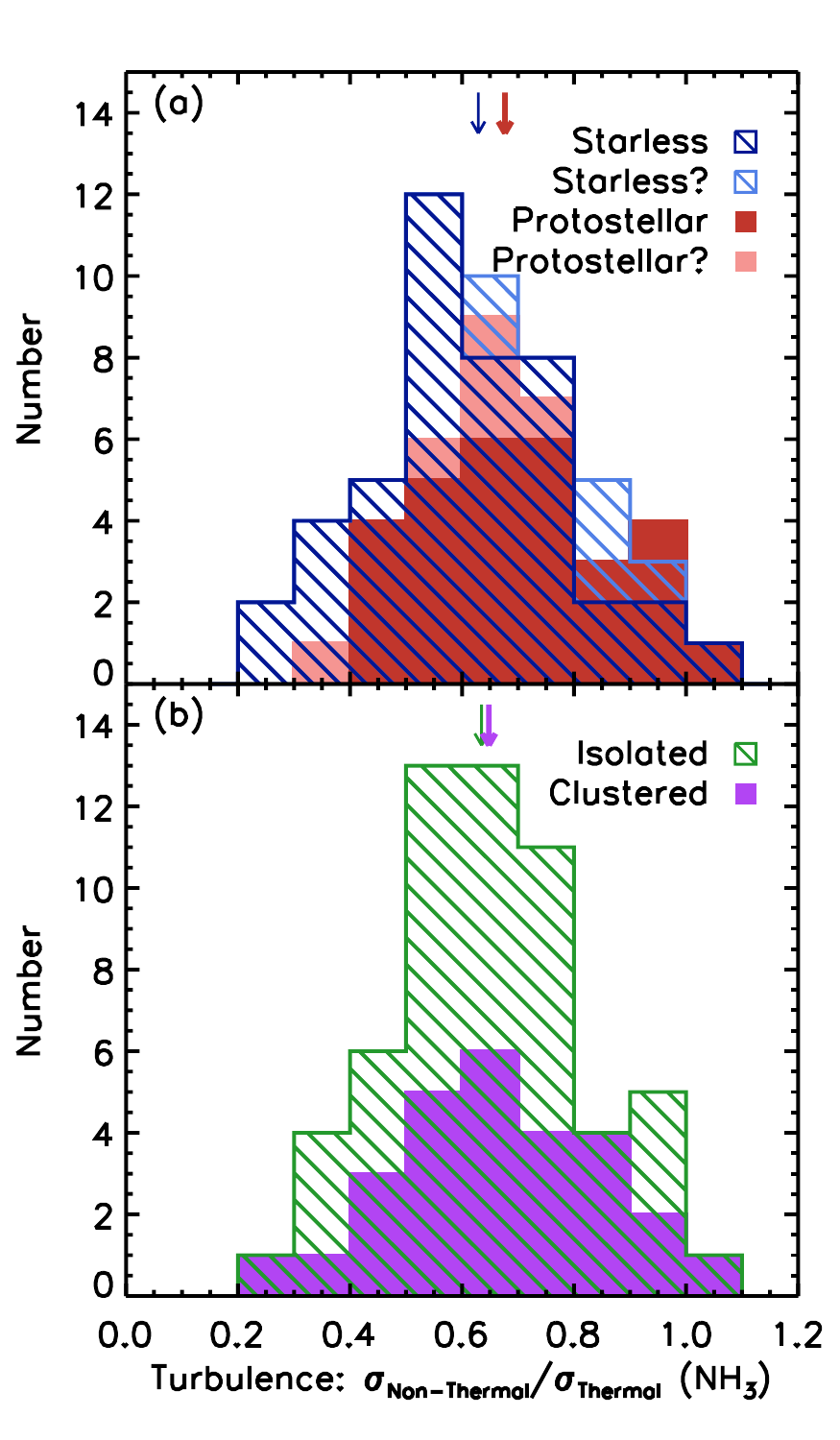}
\caption{Histograms of the ratio of non-thermal to thermal linewidths for NH$_3$ broken down by protostellar content and location within a cluster. Panel a shows starless (blue/dashed) and protostellar (red/solid) cores. Uncertain identifications are identified by lighter colors and are stacked on top of the certain identifications. Panel b shows isolated (green/dashed) cores and cores inside of clusters (purple/solid). Arrows show median values of the distributions.}
\label{Turbulence}
\end{figure}

\subsection{Linewidths}
\label{Line-widths}

The total linewidth observed can be decomposed into a portion from thermal motion and a portion from non-thermal motion. These non-thermal motions may be due to chaotic turbulence or ordered bulk motion (e.g. rotation or shear). In general, all our cores are quiescent, in the sense that the non-thermal velocity dispersion is less than the thermal component. Nonetheless, it is this non-thermal component that is most interesting for understanding the dynamics of the cores, and removing the thermal component is required for making contrasts between different populations, since otherwise the differences in temperature would dilute or disguise the results.

Our non-thermal linewidths are calculated as
\begin{equation}
\sigma_\mathrm{thermal} (\mathrm{NH_3}) = \sqrt{\frac{k  T_\mathrm{kin}}{17  m_H}} 
\end{equation}
and
\begin{equation}
\sigma_\mathrm{thermal} (\mathrm{CCS}) = \sqrt{\frac{k  T_\mathrm{kin}}{56  m_H}}
\end{equation}
where 17 and 56 are the molecular weights of NH$_3$ and CCS respectively. Our linewidths are measured in terms of the Gaussian $\sigma$, so we simply subtract out the thermal components for each species to find

\begin{equation}
\sigma_\mathrm{non-thermal}= \sqrt{\sigma^2_\mathrm{total} - \sigma^2_\mathrm{thermal}},
\end{equation}
where $\sigma_\mathrm{total}$ is the measured $\sigma$ of the two different species. Our linewidths for NH$_3$ are corrected for optical depth, but this was not possible for CCS. 

This calculation also uses the temperature calculated for the NH$_3$ line to estimate the non-thermal width of CCS. This assumption will be wrong in the case where the bulk of the CCS emission comes from a different physical region than the NH$_3$. In particular, we might expect the CCS emission to emanate from a less dense and more turbulent surrounding envelope which is less shielded and thus warmer than the very cold and dense region where NH$_3$ is found (but CCS is depleted). Indeed, a comparison of non-thermal linewidths (Figure \ref{NonThermalVsNonThermal}) reveals larger non-thermal linewidths in the CCS, suggesting that most of the emission from these two molecules arises from physically distinct regions. If the CCS comes from a warmer portion of the core, we would underestimate the thermal component of the CCS line and hence overestimate the non-thermal component. However, due to the relatively large mass of CCS, the thermal linewidth is small (0.04 km/s) at T = 10 K, and generally much smaller than the non-thermal component, so this apparent difference in non-thermal linewidths is probably due to the CCS envelope being genuinely more turbulent. 
 
 \subsubsection{Non-thermal linewidth ($\sigma_\mathrm{non-thermal}$) of NH$_3$}
 
Figure \ref{ThermalVsNon} shows the thermal and non-thermal linewidths for NH$_3$ and confirm that most of the cores in our sample are quiescent. Figure \ref{NonThermalWidths} shows non-thermal linewidths for NH$_3$ and CCS divided into the same classes as before. In comparing starless to protostellar cores, the non-thermal linewidth of NH$_3$ is larger in protostellar cores. The non-thermal linewidth for NH$_3$ is normally distributed, and stellar content is significant at the P = 4\e{-3} level. The sense is that protostellar cores have more non-thermal linewidth (i.e. are more ``turbulent") than starless cores, though the difference is only 0.025 $\pm$ 0.011 km/s, or just at the resolution limit of our survey. As turbulent cores are less likely to collapse and form stars, the increased linewidth in these cores is more likely to be an effect rather than a cause of star formation, perhaps indicating that feedback from protostars are churning up their natal cores. We further illustrate this point in Figure~\ref{Turbulence} by showing the ratio of non-thermal to thermal linewidths in NH$_3$, again divided by class. 

\citet{Caselli:1995} show that for a sample of 6 cores within Orion B, NH$_3$ linewidth decreases with distance from a cluster. Although our clustered cores are somewhat more turbulent, we do not see a statistically significant difference in non-thermal linewidths between clustered and isolated cores. This suggests that the influence of relatively weak clusters such as IC348 and NGC1333 in Perseus may be significantly weaker than the influence of massive clusters such as those in Orion.

 \subsubsection{Non-thermal linewidth ($\sigma_\mathrm{non-thermal}$) of CCS }

For the more diffuse CCS envelope the situation is somewhat less clear -- starless cores are more prevalent at both low ($\sigma < 0.15$ km/s) and high ($\sigma > 0.25$ km/s) linewidths.  Note that the x-axis scale is different in the two sides of Figure~\ref{NonThermalWidths}: the non-thermal linewidths in CCS are generally larger than in NH$_3$. This is also true on a case-by-case basis. As illustrated in Figure \ref{NonThermalWidths}, there does not seem to be a significant difference between starless and protostellar cores when it comes to how much greater the CCS non-thermal linewidth is than the NH$_3$. The non-thermal linewidth of CCS is not normally distributed, even after a logarithmic transformation. In any event, neither cluster environment nor stellar content appeared to have a significant influence. No significant difference is seen in the clustered/isolated comparison (panels c and d), although the lack of CCS detections in clusters precludes any strong statement. 

\subsection{Masses and Stability}
\label{MassStability}
The core mass distributions within Perseus have  already been studied extensively \citep{Hatchell:2008,Enoch:2008}. Our contribution of a better temperature estimate for each core adds only a small correction to the previous studies which have had to assume a temperature for each core (typically 10K for prestellar and 15K for protostellar). These authors have also distinguished between prestellar and protostellar mass distributions, but have not looked at the difference between clustered and isolated regions. Since we use T$_\mathrm{kin}$ from NH$_3$ to estimate T$_\mathrm{dust}$ in these masses, in cores where this assumption does not hold we will have mass errors in the same direction and magnitude as for N(H$_2$). See \S\ref{ColDensity} for details. 

\citet{Enoch:2008} looks at the stability of prestellar and protostellar cores in Perseus, as does \citet{Kirk:2007} using their N$_2$H$^+$ survey in combination with the SCUBA cores. We carry out much the same analysis, but with the two following modifications to the \citet{Enoch:2008} results. First, we correct all dust masses by using our NH$_3$ temperatures, so our core masses are related to those derived from the Bolocam map in \citet{Enoch:2008} via

\begin{equation}
M_\mathrm{dust} = M_\mathrm{Enoch} \frac{e^{12.85/T}-1}{e^{12.85/10}-1}.
\end{equation}
Second, we have a different division between protostellar and starless cores, as explained in \S\ref{ID}. None of these modifications substantially change the conclusions of \citet{Enoch:2008}, which are also broadly in agreement with the parallel (but independent) datasets used in the studies of \citet{Hatchell:2008} and \citet{Kirk:2007}. 

We adopt the same assumption for the radial distribution within a core as \citet{Enoch:2008}, namely a power-law density profile with $\rho \sim r^{-1.5}$ so that we calculate $\alpha = M_\mathrm{dyn}/M_\mathrm{dust}$ as defined in \citet{Bertoldi:1992} with
\begin{equation}
M_\mathrm{dyn} = \frac{4 R \sigma_\mathrm{dyn}^2}{G},
\end{equation}
where G = $1/232$ in units of parsecs, solar masses, and km/s. $\sigma_\mathrm{dyn}$ is the combination of thermal linewidth for H$_2$ and non-thermal linewidth of NH$_3$ added in quadrature. R is the radius of the Bolocam core as given in \citet{Enoch:2008}; initial indications from a GBT mapping project of some of the Bolocam cores in Perseus \citep{Pineda:2009} suggests that the extent of NH$_3$ emission is comparable to the extent of Bolocam emission, and thus that the two methods are tracing similar material. This justifies our combination of Bolocam extent and mass with NH$_3$ linewidths.

The virial estimate depends on T$_\mathrm{kin}$ in two ways. However, since our NH$_3$ spectrum provide high-quality linewidths and temperatures, we expect very small errors on our non-thermal linewidths. Thus the dominant dependence on T$_\mathrm{kin}$ is through M$_\mathrm{dust}$ in the denominator.

\begin{figure}
\includegraphics{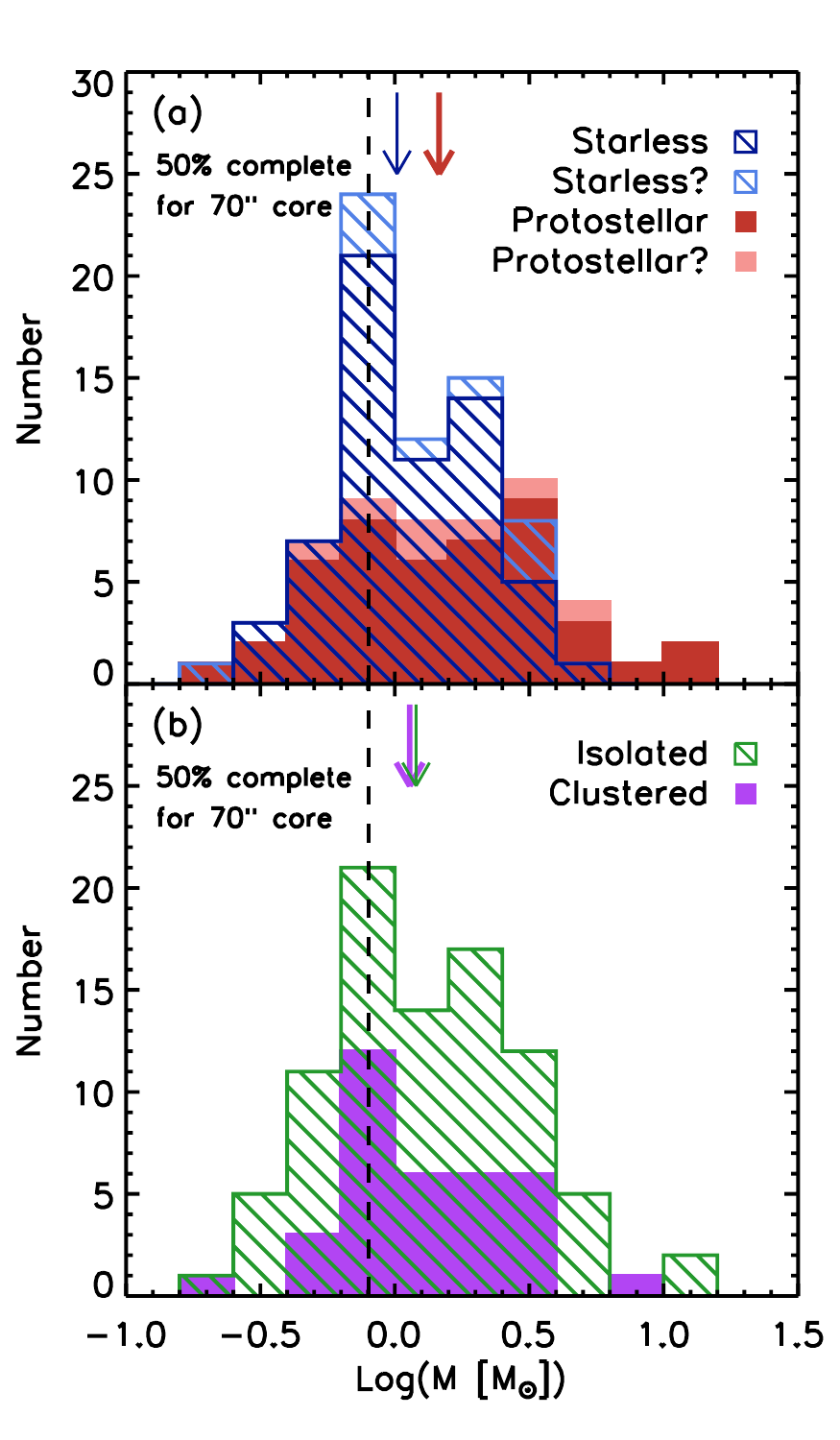}
\caption{Histograms of core masses broken down by protostellar content and location within a cluster. Panel a shows starless (blue/dashed) and protostellar (red/solid) cores. Uncertain identifications are light colors and are stacked on top of the certain identifications. Panel b shows isolated (green/dashed) cores and cores inside of clusters (purple/solid). Arrows show median values. The 50\% completeness limit is shown for the average source size (70\arcsec).}
\label{Mass}
\end{figure}

\subsubsection{Masses: M$_\mathrm{dust}$}

Figure \ref{Mass} shows the distribution of core masses for both starless/protostellar and clustered/isolated. The distribution is roughly log-normal, though likely influenced on the low-mass side by incompleteness; the 50\% completeness limit for the average source size (70\arcsec) is at 0.8 M$_{\sun}$ as estimated in \citet{Enoch:2006}. We find that cluster environment is insignificant, but protostellar cores have 1.32 times the median mass of starless cores (95\% Confidence Interval:[1.00,1.76] P = 0.05). This is consistent with what \citet{Hatchell:2008} found for cores in Perseus using the SCUBA data, where a K-S test found a 4\% chance that the masses of protostellar and starless cores were drawn from the same distribution (assuming 10K for starless and 15K for protostellar). Since NH$_3$ temperatures for protostellar cores are typically less than 15 K, \citet{Hatchell:2008} are underestimating the masses for protostellar cores relative to ours, drawing the two populations somewhat closer together. This explains why our test, despite having more degrees of freedom, provides roughly the same level of confidence that these populations are different. Levene's test reveals that the variances of the starless and protostellar distributions are unequal at the P = 0.01 level.

\subsubsection{Virial Parameter $\alpha$: M$_\mathrm{dyn}$/M$_\mathrm{dust}$}

\begin{figure}
\includegraphics{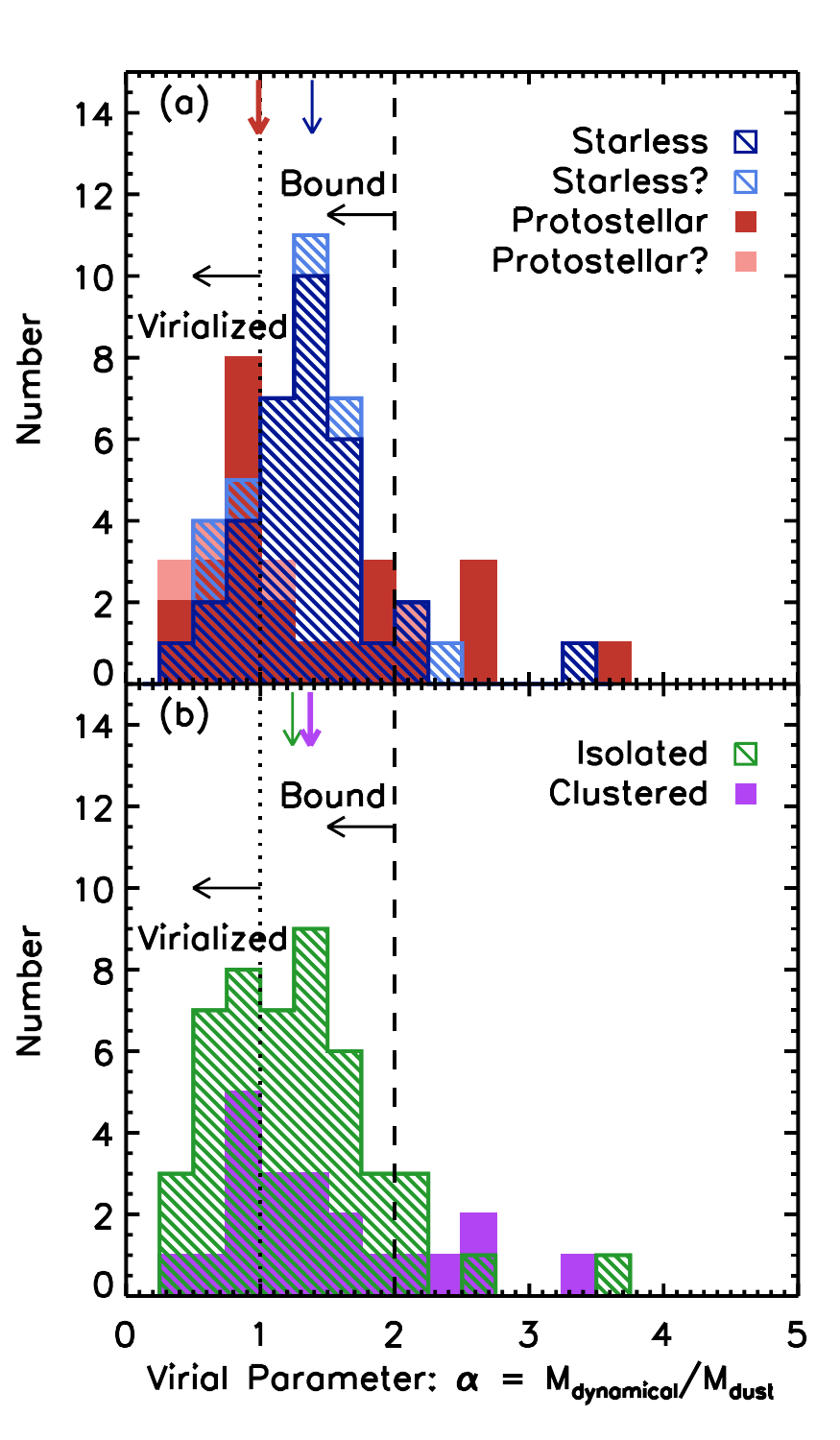}
\caption{Histograms of the virial parameter, $\alpha$. Panel a shows starless (blue/dashed) and protostellar (red/solid) cores. Uncertain identifications are lighter colors and are stacked on top of the certain identifications. Panel b shows isolated (green/dashed) cores and cores inside of clusters (purple/solid). Dashed and dotted lines show the criteria for gravitationally bound and in gravitational virial equilibrium. Vertical arrows show median values.}  
\label{Alpha}
\end{figure}

Figure \ref{Alpha} shows the distribution of the virial parameter ($\alpha$) for both starless/protostellar and clustered/isolated. Again, this variable requires a logarithmic transform to make it normal, and our test shows no significant influence from either stellar content or environment. Most of our objects are gravitationally bound ($\alpha < 2$) and roughly a third are in gravitational virial equilibrium ($\alpha < 1$). This is in agreement with \citet{Enoch:2008} (which is a similar analysis on the same datasets), but in stark contrast with \citet{Kirk:2007}, who find via a similar analysis of N$_2$H$^+$ and SCUBA cores that no cores in Perseus are in gravitational virial equilibrium and few are gravitationally bound. 

Rather than attempt a detailed reconciliation between these two studies, which rely on different algorithms to define clumps in maps with different noise and sensitivity limits, we cross-match our Bolocam sources with SCUBA positions within 60\arcsec and compare average properties of the two populations of cores. This method allows us to understand why our cores are all bound while the cores in \citet{Kirk:2007} are all unbound, but unfortunately does not answer the question of which method better determines whether a core is truly bound. The largest difference comes from the masses and sizes of the cores defined from the thermal emission maps; the non-thermal component of the N$_2$H$^+$ lines are very similar to the NH$_3$ non-thermal linewidths. Compared to our Bolocam cores, the SCUBA cores of \citet{Kirk:2007} are: much less massive (62\%) and somewhat smaller (24\%), with slightly higher total (temperature-adjusted) linewidths (110\%). Since
\begin{equation}
\alpha \sim (R \sigma^2)/M
\end{equation}
if these SCUBA cores were scaled by these median values, $\alpha$ would change by a factor of 2.7 and the results would be in rough accord with our own -- mostly bound cores with a significant number virialized. 

\section{Comparison with Other Similar Surveys}
\label{Comparison}

Apart from the other studies carried out in Perseus which we discussed above, we also compare our results to those of \citet{Jijina:1999}, long the standard compilation of NH$_3$ core properties, and the NH$_3$ and extinction study of the Pipe Nebula which was constructed in a fairly similar fashion to ours \citep{Lada:2008}.

\begin{figure*}
\includegraphics[scale=.87]{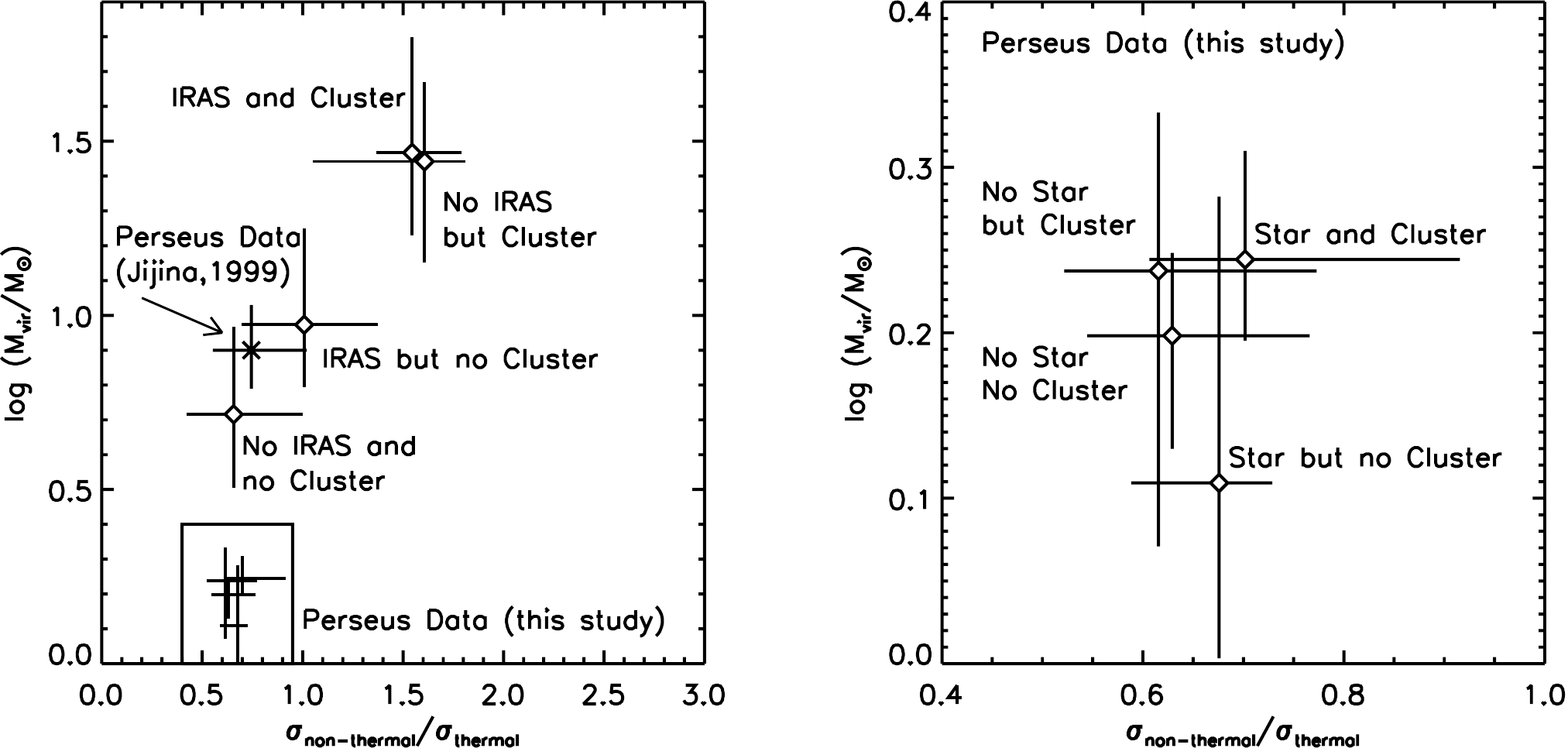}
\caption{Median and quartile log (M$_{vir}$/M$_{\sun}$) vs. $\sigma_\mathrm{non-thermal}/\sigma_\mathrm{thermal}$ for the four different sub-samples of objects as in Figure 4 from \citet{Jijina:1999}. On the left, the \citet{Jijina:1999} results (IRAS point sources were used as a proxy for protostellar), which span a much larger range in both axes and exhibit more separation between classifications than our Perseus data. Cores in Perseus from the \citet{Jijina:1999} database are typically more massive. We zoom into our data on the right, illustrating that our different classes of objects largely overlap in this diagram.}
\label{JijinaSmall}
\end{figure*}

We provide a comparison with just one of the results from \citet{Jijina:1999} which illustrates some of the advantages our uniform survey provides. \citet{Jijina:1999} find that cores within clusters are more massive (as estimated from their NH$_3$ lines assuming virial equilibrium) and have a higher ratio of non-thermal to thermal linewidths. By comparison, all our objects fit within a narrow range in Figure 4 from \citet{Jijina:1999} (see Figure~\ref{JijinaSmall}), and there does not appear to be significant distinctions between the sub-samples. The \citet{Jijina:1999} sub-sample medians run from log (M$_{vir}$/M$_{\sun}$) = 0.7 to 1.5, and  $\sigma_\mathrm{non-thermal}/\sigma_\mathrm{thermal}$ runs from from 0.6 to 1.6. As shown in Figure \ref{JijinaSmall}, all our objects are clustered near the lowest mass (log (M$_{vir}$/M$_{\sun}$) between 0.1 and 0.4) and least turbulent objects ($\sigma_\mathrm{non-thermal}/\sigma_\mathrm{thermal}$ between 0.5 and 0.8). 

The \citet{Jijina:1999} sample includes cores from Perseus, most of which (15/20) are considered to be protostellar and isolated. Our objects within Perseus have similar levels of turbulence, but are significantly less massive than those of \citet{Jijina:1999}. Two simple explanations for this are that \citet{Jijina:1999} assume a distance of 350 pc for Perseus, where we use 250 pc, and that the observations of Perseus in this other sample were taken largely from Effelsberg (40\arcsec  beam) and Haystack (88\arcsec  beam), the latter of which in particular is much larger than our 31\arcsec  beam, and thus might easily blend two sources which we resolve. Both these effects would tend to increase the masses. The much larger range in core properties identified in the full sample of \citet{Jijina:1999} could be the result of cores being influenced by larger scale (i.e. cloud) properties.

\citet{Lada:2008} conclude that for the Pipe Nebula, a nearby molecular cloud with little star formation, most cores are thermally-dominated objects bound only by external pressure. This contrasts with our cores, which appear to be gravitationally bound. The Pipe core masses come from an extinction map, and their dynamical information comes mainly from a pointed survey of C$^{18}$O (objects were examined for NH$_3$ but with a low success rate). Due to the lack of NH$_3$ detections, \citet{Lada:2008} assumed 10 K for every core. 

It is possible that the objects detected in the Pipe are fundamentally different objects than the millimeter cores studied herein. \citet{Alves:2007} note that the cores identified within the Pipe have typical densities between 5\e{3} cm$^{-3}$ and 2\e{4} cm$^{-3}$, much lower than the densities of the Bolocam cores discussed herein.  The low rate of NH$_3$ detections in the study of \citet{Rathborne:2008} appears to confirm that these cores are relatively low density. The Pipe cores could be either younger cores (which will eventually become denser) or a distinctly different population of objects which would be undetected by millimeter surveys at the distance of Perseus. In the latter scenario, the cores we define in Perseus represent just a portion of the true core mass function.

\citet{Lada:2008} do a least-squares fit of Log(Mass) versus Log($\alpha$) and find a slope of -2/3, in agreement with the prediction of \citet{Bertoldi:1992} for pressure-confined cores. We perform a similar experiment on our sample of objects. Our masses span a similar range, but in contrast to \citet{Lada:2008}, almost all of our cores are consistent with being gravitationally bound. This could be due to genuine differences between the two regions under consideration as discussed above or due to the different linewidths of the tracers used. Linewidths are the obvious source of systematic differences simply because of dominance of systematic uncertainty of this term ($\alpha \sim \sigma_\mathrm{dyn}^2$). C$^{18}$O traces a lower density envelope than NH$_3$, and thus using it to estimate the linewidths of cores will produce an overestimate of the non-thermal component. This overestimate was on average a factor of 4 within the Pipe cores observed in both tracers.

An unweighted least-squares fit to our full data set provides the same slope (M $\sim \alpha ^ {-0.65}$), however a closer look reveals some cautions with this analysis. First, we calculate the uncertainty on our derived parameters, propagating the errors on masses and radii quoted in \citet{Enoch:2006} and the errors on temperatures and linewidths from \citet{Rosolowsky:2008} into our derived quantities. This ignores systematic errors in the mass estimate. With these estimates we are able to state that the low-mass cores are almost all consistent with being gravitationally bound without external pressure, and that high-mass cores are unstable to gravitational collapse (unless supported by magnetic fields).

To consider the influence of selection effects, we take the completeness limits on Bolocam cores from \citet{Enoch:2006}. One constraint is simply due to the beam size, which prohibits detection of objects with FWHM $<$ 34\arcsec. There are also curves in mass and radius derived from Monte Carlo experiments on simulated data and simulated reduction routines which are roughly flux limits (high mass objects are difficult to detect if too extended) -- we use the 90\% completeness limit presented. These curves are transformed into $\alpha$ and mass constraints which requires an estimate of $\sigma_\mathrm{non-thermal}$.  Our ability to measure the linewidth is generally not a constraint within our survey, but since our completeness in alpha is thus dependent on three variables (mass, radius, and linewidth), it is not possible to represent it faithfully in this plot. 

\begin{figure*}
\includegraphics[scale=1]{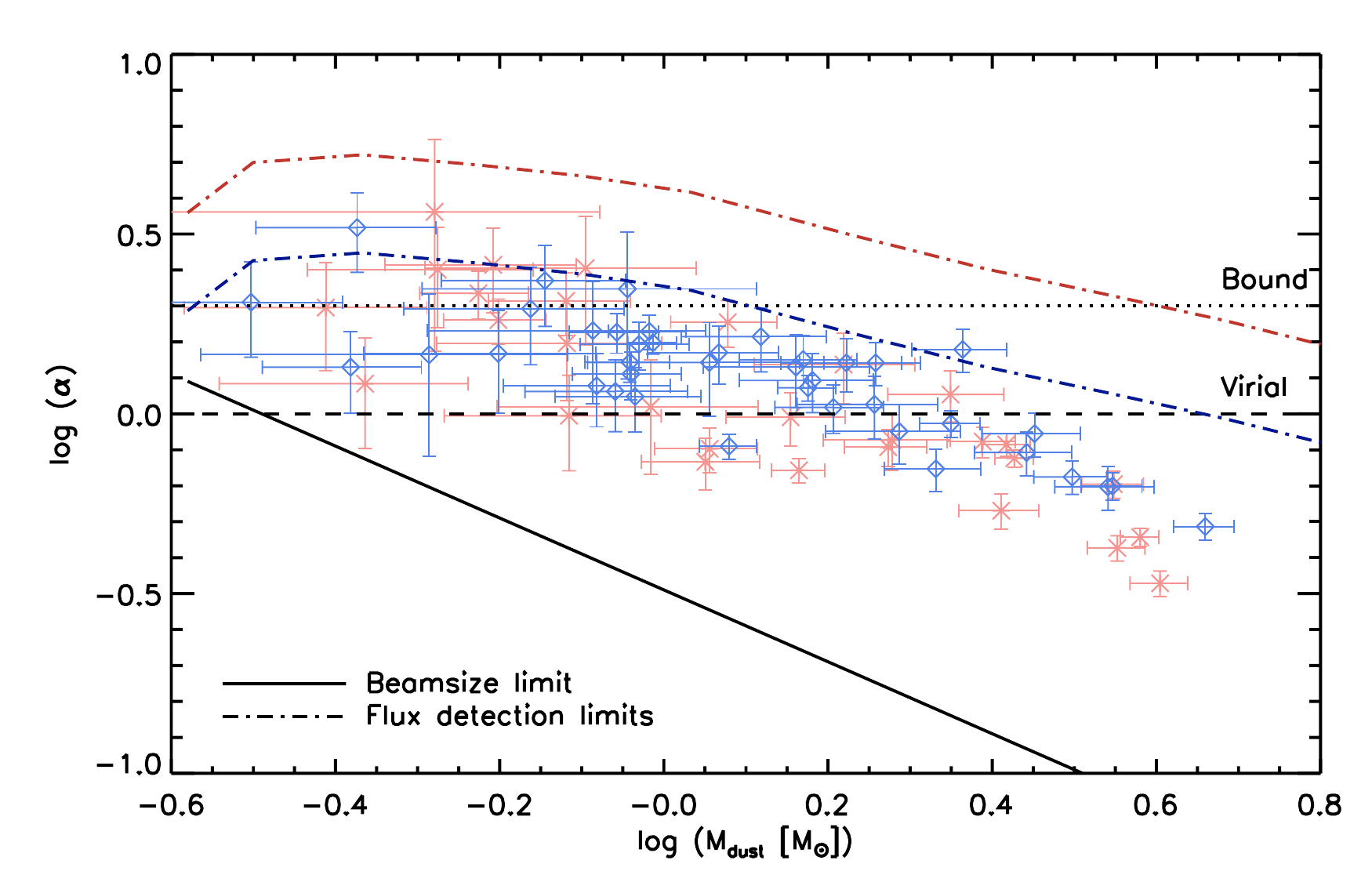}
\caption{Log masses and virial parameter $\alpha$ for our sample of cores broken down into starless (blue/open diamonds) and protostellar (red/stars). The horizontal broken black lines show the conditions for gravitationally bound (dotted) and virial equilibrium (dashed). Colored lines show estimated completeness limits using the 75\% quartile linewidths for these two samples at the 90\% (dot-dashed) flux-detection limit. The solid black line shows an estimate of the influence of beam size in detecting compact sources. Completeness could be significantly influencing this relation.}  
\label{StarlessFull_MvsAlpha}
\end{figure*}

To provide a useful guide, we use simple upper and lower bounds for linewidths. The flux limit provides an upper bound on alpha (we can't detect things too extended/far-from-bound), so we use the 75\% quartile of the non-thermal linewidths for the protostellar (0.174 km/s) and starless (0.148 km/s) sub-samples when calculating the total linewidths of an object. This provides a reasonably conservative number upper bound on the linewidths and thus alpha. The beam-width limit is a lower bound on alpha (we can't measure a radius less than the beam), so we use the 25\% quartile linewidths for the full population. Figure \ref{StarlessFull_MvsAlpha} shows the data points, errors, and completeness estimates for starless and protostellar cores. Our completeness estimates nearly bound the data points on the upper side, and could contribute to the shape of this relation.

In particular, completeness effects could be severely limiting our ability to detect self-gravitating cores at masses below 1 $M_{\odot}$. At high-masses, the constraints are not as severe. That is, we can be reasonably confident that there are no high-mass cores which are gravitationally unbound, and the highest mass cores are all estimated to be significantly self-gravitating (i.e. likely undergoing collapse).

\section {Conclusion}

As summarized in Table~\ref{ANOVAtable}, the median values of cores in clusters are statistically significantly different from the median values of isolated cores in the following ways:
\begin{enumerate}
\item Higher kinetic temperatures (T$_\mathrm{kin}$) \\
	Cluster Cores = 12.9 K\\
	Isolated Cores = 10.8 K
\item Lower fractional abundances of CCS \\
	Cluster Cores = 0.6\e{-9} \\
	Isolated Cores = 2.0\e{-9} 
\item Lower fractional abundances of NH$_3$\\
	Cluster Cores = 1.2\e{-8} \\
	Isolated Cores = 2.9\e{-8}
\end{enumerate}

The median values of cores associated with protostars are statistically significantly different from the median values of starless cores in the following ways:
\begin{enumerate}
\item Higher kinetic temperatures (T$_\mathrm{kin}$) \\
	Protostellar = 11.9 K\\
	Starless = 10.6 K
\item Higher excitation temperatures (T$_\mathrm{ex}$) \\
	Protostellar = 7.4 K \\
	Starless = 6.1 K
\item Higher column density (N(H$_2$))\\
	Protostellar = 1.2\e{22} cm$^{-2}$ \\
	Starless = 0.6\e{22} cm$^{-2}$
\item Lower fractional abundance of CCS \\
	Protostellar = 1.4\e{-9} \\
	Starless =  2.4\e{-9}
\item Higher non-thermal motion ($\sigma_\mathrm{non-thermal}$ NH$_3$)\\
	Protostellar = 0.14 km/s\\
	Starless = 0.11 km/s
\item Higher masses \\ 
	Protostellar =1.5 M$_{\odot}$\\
	Starless = 1.0 M$_{\odot}$
\end{enumerate}
Neither cluster environment nor protostellar content makes a significant difference to the dynamical state of cores as estimated by the virial parameter. Most cores are bound.

We are not able to assess whether differences in core properties arise from the influence of protostars/clusters or whether protostars and clusters form around cores with a given set of properties. Nonetheless, we propose explanations for these trends which present a coherent picture. 

We see evidence that the cluster environment has a significant influence on the $\sim$ 10$^4$ cm$^{-3}$ gas characteristic of NH$_3$. This is most straightforwardly interpreted as the influence of radiation, which could (1) increase the temperature and modify the chemistry, decreasing the abundance of both  (2) CCS and (3) NH$_3$. This sort of behavior is seen, for example, in the models of \citet{Lee:2004} in which CS (which behaves like CCS) and NH$_3$ abundances are suppressed in the outer regions of a core by the presence of a UV field. The slightly higher column density seen in clustered cores could be responsible for lowering the CCS abundance (via depletion), but can not easily explain the lowered abundance of NH$_3$.

The small temperature increase seen is not sufficient to be distinguishable in the virial analysis we have carried out. We see no evidence that the mass distribution of cores is different between clusters and isolated regions, although further study of this question in warranted. This suggests that low mass clusters may form most of their stars in a mode fundamentally similar to isolated star formation. Other forms of feedback, such as outflows or shells may be important in determining the properties of cores in clusters, but we are unable to point to any specific evidence of such an influence. 

Clusters with higher-mass star formation may present a better chance to study the effects of feedback and examine the differences between low- and high- mass star formation. Although the relatively low-mass stars in NGC 1333 are likely destroying the clouds \citep[see][and references therein]{Gutermuth:2008}, these mid-B stars still have a relatively weak impact on their environs compared to the O-stars which dominate a region such as Orion. Furthermore, the clusters in Perseus have between 100 and 300 stars, while the Orion Nebula Cluster has $\sim$ 1500 \citep{Porras:2003}. Studies on the ammonia clumps in Orion already suggest that NH$_3$ cores here have different properties from these seen in Perseus \citep{Harju:1993,Caselli:1995,Wilson:1999}. The cores in Orion from \citet{Jijina:1999} show a large range in many variables, with some clear distinctions (T$_\mathrm{kin}$ is greater in clusters than in isolated cores), but larger samples would be useful. A study which contained a large number of NH$_3$ cores both near and far from the clusters in Orion would provide definitive statements about the influence of massive clusters on dense cores. 

We propose that some of the properties of protostellar cores are due to the presence of a protostar, while other properties are indicative of the environment necessary for star formation. Higher non-thermal linewidths and temperatures would seem to inhibit rather than promote star formation. Since we see both these features in protostellar cores we propose that we are seeing the influence of the protostar on its natal core. Chemical changes would seem likely to be either caused by star formation or due to an independent variable such as time (since it takes time for cores to both evolve chemically and to collapse into a star). 

On the other hand, it has been suggested that protostars only form in cores with sufficiently high column density \citet{Johnstone:2004}. We interpret our higher excitation temperatures as arising from higher physical densities, which might promote star formation \citep[for instance by changing the thermal physics as in][]{Larson:2005}. Under the competitive accretion model \citep{Bonnell:2001}, cores which are already going to form a star might be expected to draw in additional mass, but it is also plausible that more massive cores are simply more likely to collapse. Thus we are left with the conclusion that some of the differences in protostellar cores are due to the influence of those young stars, while other differences may explain why those particular cores are chosen out of the ensemble to birth a star. Our lack of distinction between starless and protostellar cores on the question of dynamical stability seems to suggest that most of our starless cores are relatively long-lived rather than transient structures.

Ultimately, a theory of star formation should be able to reproduce the properties of the cores presented here, both pre- and post- the star formation event, both inside clusters and in isolated regions. Matching all these properties will provide much tighter constraints on the physics of star formation than simply matching the mass function of these cores. Much progress has already been made by numerical simulations of star formation within a molecular cloud. To match the observations herein, these models must include realistic thermal physics and should include at least a simple chemical model in order to provide an accurate and unbiased comparison with the physical conditions of the gas traced by NH$_3$ and CCS.

\acknowledgments

The Green Bank Telescope is operated by the National Radio Astronomy Observatory. The National Radio Astronomy Observatory is a facility of the National Science Foundation, operated under cooperative agreement by Associated Universities, Inc. J. E. P. and J. B. F. are supported by a generous grant from the NRAO Student Observing Support Program (GSSP06-0015). The work of E. R. is supported by an NSF Astronomy and Astrophysics Postdoctoral Fellowship (AST-0502605). We thank our anonymous referee who helped to improve this paper.

\bibliographystyle{apj}
\bibliography{NH3}

\clearpage

\section{Appendix A: Classification of Cores as Starless or Protostellar}

In principle, the question we want to ask when classifying a dense core as protostellar or starless is: ``Is a star which formed within this dense core still embedded within the core?" Evolution presents an intractable complication to answering this question. A dense core may survive the formation of a star. This star is then subject to different forces (it does not feel gas pressure, for instance) and may move away from the core. At some point we will no longer be able to associate this YSO with its natal core. This problem may be particularly intense in clusters where we must also worry about another problem, that a neighboring star may pass close to a different dense core and thus appear to be associated. Of course, if we wish to measure the generic impact of a young star upon a core, associating this star with this new core may still be acceptable.

Even putting aside that problem, dense cores form part of a smooth larger distribution of material and thus may have many legitimate different definitions of their extent. There are thus different valid definitions of what constitutes an embedded YSO. This is complicated by the fact that our analysis depends on two different data sets, a millimeter continuum map for which a measure of the 2-dimensional extent may be made, and pointed NH$_3$ observations, for which no such measure can be established. Indeed, the situation is worse, for some of our NH$_3$ pointings are at positions for which no sensible extent can be measured from the millimeter map, either because the pointing is outside the Bolocam map, or because the millimeter continuum at a position has too low a signal-to-noise ratio. To maintain our sample size and limit additional biases we seek a general metric for whether the NH$_3$ gas observed at a given position is associated with recent star formation or not.

We combine the results of two previous studies which have made this distinction with two of our own methods. The two literature studies have more sophisticated methods, but rely on data not available for our entire survey.

\citet{Enoch:2008} provides a classification of Bolocam cores as starless or protostellar based on constructing a catalog of cold YSOs. This catalog is a subset of the candidate young stellar objects (YSOc) from the c2d catalog (\citet{Evans:2007}) with some additional very red objects included. Then, using the sizes for Bolocam cores defined in \citet{Enoch:2006}, they consider a core to be protostellar if there is a cold YSO within 1 FWHM of the center of the core. This method provides a reasonably clean definition for millimeter cores, but it is not extensible to objects for which we are unable to define a millimeter core. Furthermore, some of the cores in \citet{Enoch:2006} are quite large (mean minor-axis FWHM is 58\arcsec $\pm$ 17\arcsec and mean major-axis FWHM is 80\arcsec $\pm$ 27\arcsec, with major axes FWHM extending out to 150\arcsec). Our GBT beam is probing gas within about 30\arcsec of the center of these cores, so while a YSO may legitimately be impacting the total flux from a large millimeter core, it may not have any impact on the dense gas we are tracing.

\citet{Hatchell:2007} also builds a catalog of young YSOs (class 0 or 1) by considering SCUBA-identified cores (supplementing this list with additional Bolocam cores) and constructing and fitting SEDs from these long-wavelength fluxes with many additional shorter wavelengths including 2MASS and IRAC fluxes. The association of shorter wavelength fluxes with the (sub)millimeter points and the inclusion of short wavelength (2MASS) points in constructing the SED are two essential differences with the \citet{Enoch:2008} results. The two methods agree reasonably well, but disagree on 10 out of the 76 cores which both classify. We consider every core where these methods disagree to be an uncertain classification.

Our own two methods are rougher but applicable to all sources. First, we take the the cold YSO list of \citet{Enoch:2008} and the YSOc list from the c2d catalog (\citet{Evans:2007}) and at each GBT pointing we see whether one of these objects falls within 1 FWHM of the beam (30\arcsec). This method shows good agreement with the two classifications from the literature, disagreeing with one of these methods only 13 times (and 10 of these times the two literature methods disagree, as noted above). 

Second, we examine the proximity and strength of 24 $\micron$ sources from the c2d MIPS data for Perseus \citep{Rebull:2007}. Some 24 $\micron$ sources are background galaxies, but the majority are YSOs. By determining the average spatial density of 24$\micron$ sources as a function of intensity we are able to make an estimate for each position: ``Is there a 24$\micron$ source close and bright enough that it is unlikely to be there by chance?'' Not including the ambiguous cases discussed above, this method categorizes 43 more cores as being protostellar than the other methods, largely due to the fact that cores are very clustered, which places many starless cores in close proximity to a bright 24 $\micron$ source which nonetheless lives in another core. There is also significant bright, diffuse 24 $\micron$ emission towards the east of Perseus.

The 24 $\micron$ method was very useful in assessing positions where neither literature method provides a classification. Since these are necessarily weaker cores (or outside the area mapped), it is not surprising that they are overwhelmingly starless. A final case of note is the mini-cluster L1455, which lies on the edge of the Bolocam and c2d surveys. Here we were restricted to the 24 $\micron$ criteria and other literature information about the YSOs in this well-studied region. There were 4 cores considered in this fashion. We did use the uniform 24 $\micron$ criteria to divide cores into two categories ``influenced'' and ``not-influenced'' and produced the plots within this paper under this division, though we do not show these plots. In all cases the same qualitative behavior was observed, in the sense that starless cores are mostly ``not-influenced'' and protostellar cores are mostly ``influenced''. We mention the one case where this distinction produces substantially different distributions within the text (fractional abundance of CCS, \S\ref{FracAbundanceOfCCS}).

In summary, if any two of the three main methods disagree we examine the images my hand and consider the reasons for the disagreement. This is normally due to a dispute over what the extent of the core really is, or whether a somewhat distant YSO should be considered associated, but sometimes hinges on the quality of the SED used to make the classification. We make an assessment and assign this object a classification, but mark it as uncertain. These objects are all presented with a brief comment in Table~\ref{IDs}.

\clearpage
\LongTables
\begin{deluxetable}{ccccccccp{2.5in}}
\tabletypesize{\scriptsize}
\tablecaption{Classification of objects as Protostellar/Starless}
\tablehead{NH3SRC\tablenotemark{a} & NH$_3$ & Bolocam & Class\tablenotemark{b} &	Enoch\tablenotemark{c} &	Hatchell\tablenotemark{d} & c2d YSOc\tablenotemark{e} & 	24 $\micron$&	Comments\\
\#& detected? & core?& & & &in beam?& &
}
\startdata
1	& N	&	N &	L &	\nodata	&	\nodata	&	N	&	N	&	Outside of Bolocam and SCUBA.\\ 
2	& Y	&	N &	L &	N	&	\nodata	&	N	&	N	&	\\ 
3	& Y	&	Y &	L &	N	&	\nodata	&	N	&	N	&	\\ 
4	& Y	&	Y &	L &	N	&	\nodata	&	N	&	N	&	\\ 
5	& Y	&	Y &	L &	N	&	\nodata	&	N	&	Y	&	\\ 
6	& Y	&	Y &	L &	N	&	\nodata	&	N	&	N	&	Outside of SCUBA.\\ 
7	& Y	&	Y &	S &	Y	&	Y	&	Y	&	Y	&	\\ 
8	& Y	&	N &	S?&	\nodata	&	Y	&	N	&	Y	&	Hatchell lists a YSO, but close to NH3SRC 7.\\ 
9	& Y	&	Y &	L &	N	&	\nodata	&	N	&	N	&	Outside of SCUBA.\\ 
10	& Y	&	N &	L&	\nodata	&	\nodata	&	N	&	Y	&	\\ 
11	& Y	&	Y &	L &	N	&	\nodata	&	N	&	N	&	Outside of SCUBA.\\ 
12	& Y	&	Y &	S &	Y	&	Y	&	Y	&	Y	&	\\ 
13	& Y	&	Y &	L &	N	&	\nodata	&	N	&	N	&	Outside of SCUBA.\\ 
14	& Y	&	Y &	S &	Y	&	Y	&	Y	&	Y	&	\\ 
15	& Y	&	Y &	L &	N	&	N	&	N	&	Y	&	\\ 
16	& Y	&	Y &	L &	N	&	\nodata	&	N	&	N	&	Outside of SCUBA.\\ 
17	& Y	&	Y &	L &	N	&	N	&	N	&	N	&	\\ 
18	& Y	&	Y &	L &	N	&	\nodata	&	N	&	N	&	\\ 
19	& Y	&	Y &	L &	N	&	\nodata	&	N	&	N	&	\\ 
20	& Y	&	Y &	L &	N	&	\nodata	&	N	&	N	&	\\ 
21	& Y	&	Y &	L &	N	&	\nodata	&	N	&	N	&	\\ 
22	& Y	&	Y &	S &	Y	&	Y	&	Y	&	Y	&	\\ 
23	& N	&	N &	L &	\nodata	&	\nodata	&	N	&	N	&	Outside of Bolocam and SCUBA.\\ 
24	& Y	&	Y &	L &	N	&	\nodata	&	N	&	N	&	\\ 
25	& Y	&	N &	L &	\nodata	&	\nodata	&	N	&	N	&	Outside of SCUBA.\\ 
26	& Y	&	N &	L &	\nodata	&	\nodata	&	N	&	N	&	Outside of SCUBA\\ 
27	& Y	&	N &	L &	\nodata	&	\nodata	&	N	&	N	&	Outside of Bolocam and SCUBA.\\ 
28	& Y	&	N &	L &	\nodata	&	\nodata	&	N	&	N	&	Outside of Bolocam and SCUBA.\\ 
29	& Y	&	Y &	L &	N	&	\nodata	&	N	&	N	&	\\ 
30	& Y	&	N &	L &	\nodata	&	\nodata	&	N	&	N	&\\ 
31	& Y	&	Y &	S &	Y	&	Y	&	Y	&	Y	&	\\ 
32	& Y	&	Y &	S &	Y	&	Y	&	Y	&	Y	&	\\ 
33	& Y	&	N &	L &	\nodata	&	N	&	N	&	Y	&	\\ 
34	& Y	&	Y &	S &	Y	&	Y	&	Y	&	Y	&	\\ 
35	& Y	&	Y &	S &	Y	&	Y	&	Y	&	Y	&	\\ 
36	& Y	&	N &	L&	\nodata	&	\nodata	&	N	&	Y	&	At edge of SCUBA and Bolocam. Starless in literature.\tablenotemark{f} 24 $\micron$ source in NH3SRC37. \\ 
37	& Y	&	N &	S &	\nodata	&	\nodata	&	Y	&	Y	&	\\ 
38	& Y	&	N &	L &	\nodata	&	\nodata	&	N	&	N	&	Outside of Bolocam and SCUBA. Starless in literature.\tablenotemark{f}\\ 
39	& N	&	N &	L &	\nodata	&	\nodata	&	N	&	N	&\\ 
40	& Y	&	Y &	S &	Y	&	Y	&	Y	&	Y	&	\\ 
41	& Y	&	Y &	L &	N	&	N	&	N	&	N	&	\\ 
42	& Y	&	Y &	S?&	Y	&	N	&	N	&	Y	&	Hatchell has a starless core slightly to the north but Enoch and c2d have a source on edge of GBT beam.\\ 
43	& Y	&	Y &	S &	Y	&	Y	&	Y	&	Y	&	\\ 
44	& Y	&	Y &	S &	Y	&	Y	&	Y	&	Y	&	\\ 
45	& N	&	N &	L &	\nodata	&	N	&	N	&	N	&	\\ 
46	& Y	&	Y &	S &	Y	&	Y	&	Y	&	Y	&	\\ 
47	& Y	&	N &	L?&	\nodata	&	N	&	Y	&	Y	& c2d source just within beam.\\ 
48	& Y	&	Y &	S &	Y	&	Y	&	Y	&	Y	&\\ 
49	& Y	&	Y &	L &	N	&	\nodata	&	N	&	Y	&	\\ 
50	& Y	&	Y &	S?&	N	&	Y	&	N	&	Y	&	Small Bolocam core with Enoch just outside beam. Hatchell has YSO.\\ 
51	& Y	&	Y &	L &	N	&	\nodata	&	N	&	Y	&	\\ 
52	& Y	&	Y &	S &	Y	&	\nodata	&	Y	&	Y	&	\\ 
53	& Y	&	Y &	S?&	Y	&	\nodata	&	N	&	Y	&	Big Bolocam core. Nearby c2d source outside beam.\\ 
54	& Y	&	N &	L &	\nodata	&	\nodata	&	N	&	N	&\\ 
55	& Y	&	N &	L &	\nodata	&	\nodata	&	N	&	N	&\\ 
56	& Y	&	Y &	S &	Y	&	\nodata	&	Y	&	Y	&	Multiple  c2d sources within beam and Bolocam core.\\ 
57	& Y	&	N &	L &	\nodata	&	\nodata	&	N	&	Y	&	\\ 
58	& Y	&	Y &	S &	Y	&	Y	&	Y	&	Y	&	\\ 
59	& Y	&	Y &	S &	Y	&	Y	&	Y	&	Y	&	\\ 
60	& Y	&	N &	L &	\nodata	&	\nodata	&	N	&	N	&	Outside of Bolocam and SCUBA.\\ 
61	& Y	&	N &	L?&	\nodata	&	\nodata	&	N	&	Y	& Enoch object just outside beam.\\ 
62	& Y	&	N &	L &	\nodata	&	\nodata	&	N	&	Y	&	\\ 
63	& Y	&	N &	L &	\nodata	&	\nodata	&	N	&	N	&	\\ 
64	& Y	&	Y &	S &	Y	&	Y	&	Y	&	Y	&	\\ 
65	& Y	&	Y &	S &	Y	&	Y	&	Y	&	Y	&	\\ 
66	& Y	&	Y &	S &	Y	&	Y	&	Y	&	Y	&	\\ 
67	& Y	&	Y &	S &	Y	&	Y	&	Y	&	Y	&	\\ 
68	& Y	&	N &	S &	\nodata	&	Y	&	Y	&	Y	&	\\ 
69	& Y	&	Y &	L &	N	&	N	&	N	&	Y	&	\\ 
70	& Y	&	N &	L?&	\nodata	&	Y	&	N	&	Y	&	Nearby Enoch sources are inside other cores. These fluxes probably contribute to Hatchell classification.\\ 
71	& Y	&	Y &	S &	Y	&	Y	&	Y	&	Y	&	\\ 
72	& Y	&	Y &	L?&	N	&	Y	&	N	&	Y	&	Hatchell source (\#62) has an oddly rising SED. No c2d source within beam.\\ 
73	& Y	&	Y &	L? &	Y	&	N	&	N	&	Y	&	Large Bolocam core but all c2d sources are outside of beam and inside other cores.\\ 
74	& N	&	N &	L &	\nodata	&	\nodata	&	N	&	N	&	\\ 
75	& Y	&	Y &	S &	Y	&	Y	&	Y	&	Y	&	\\ 
76	& Y	&	Y &	S &	Y	&	Y	&	Y	&	Y	&	\\ 
77	& Y	&	Y &	S &	Y	&	Y	&	Y	&	Y	&	\\ 
78	& Y	&	N &	S &	\nodata	&	Y	&	Y	&	Y	& \\ 
79	& Y	&	Y &	L?&	N	&	Y	&	N	&	Y	&	Hatchell source (\#70) has flat/rising IRAC fluxes. Not too odd, but no c2d source within beam.\\ 
80	& Y	&	Y &	L &	N	&	N	&	N	&	Y	&	\\ 
81	& Y	&	Y &	S &	Y	&	Y	&	Y	&	Y	&	\\ 
82	& Y	&	Y &	S?&	Y	&	N	&	Y	&	Y	&	Closest Enoch source in neighboring core, but c2d source at center of beam.\\ 
83	& Y	&	Y &	L &	N	&	N	&	N	&	N	&	\\ 
84	& Y	&	Y &	S &	Y	&	Y	&	Y	&	Y	&	\\ 
85	& Y	&	N &	L &	\nodata	&	\nodata	&	N	&	Y	&	\\ 
86	& Y	&	Y &	L &	N	&	N	&	N	&	N	&	\\ 
87	& Y	&	Y &	S &	Y	&	Y	&	Y	&	Y	&	\\ 
88	& Y	&	Y &	L &	N	&	N	&	N	&	Y	&	\\ 
89	& Y	&	Y &	S &	Y	&	\nodata	&	Y	&	Y	&	Outside of SCUBA.\\ 
90	& Y	&	N &	L &	\nodata	&	\nodata	&	N	&	N	&	Outside of SCUBA.\\ 
91	& Y	&	Y &	S &	Y	&	Y	&	Y	&	Y	&	\\ 
92	& N	&	N &	L &	\nodata	&	\nodata	&	N	&	N	&	\\ 
93	& Y	&	Y &	L &	N	&	\nodata	&	N	&	Y	&	\\ 
94	& N	&	N &	L &	\nodata	&	\nodata	&	N	&	N	&	\\ 
95	& Y	&	Y &	S &	Y	&	Y	&	Y	&	Y	&	\\ 
96	& Y	&	Y &	L &	N	&	\nodata	&	N	&	N	&	\\ 
97	& Y	&	Y &	L &	N	&	N	&	N	&	N	&	\\ 
98	& Y	&	N &	L &	\nodata	&	\nodata	&	N	&	N	&	\\ 
99	& Y	&	Y &	S &	Y	&	Y	&	Y	&	Y	&	\\ 
100	& N	&	N &	L &	\nodata	&	\nodata	&	N	&	N	&	Outside of Bolocam and SCUBA.\\ 
101	& Y	&	N &	L &	\nodata	&	\nodata	&	N	&	N	&	\\ 
102	& Y	&	N &	L &	\nodata	&	\nodata	&	N	&	N	&	\\ 
103	& Y	&	Y &	S &	Y	&	Y	&	Y	&	Y	&	\\ 
104	& Y	&	Y &	L &	N	&	N	&	N	&	N	&	\\ 
105	& Y	&	Y &	S &	Y	&	\nodata	&	Y	&	Y	&	\\ 
106	& N	&	N &	L &	\nodata	&	\nodata	&	N	&	N	&	\\ 
107	& Y	&	Y &	L &	N	&	\nodata	&	N	&	N	&	\\ 
108	& Y	&	Y &	L &	N	&	N	&	N	&	Y	&	\\ 
109	& Y	&	Y &	L &	N	&	\nodata	&	N	&	N	&	\\ 
110	& Y	&	N &	L &	\nodata	&	\nodata	&	N	&	N	&	\\ 
111	& Y	&	Y &	L &	N	&	\nodata	&	N	&	N	&	\\ 
112	& Y	&	Y &	L &	N	&	\nodata	&	N	&	N	&	\\ 
113	& Y	&	Y &	L &	N	&	N	&	N	&	Y	&	\\ 
114	& Y	&	Y &	L &	N	&	\nodata	&	N	&	Y	&	\\ 
115	& Y	&	N &	L &	\nodata	&	\nodata	&	N	&	Y	&	\\ 
116	& Y	&	Y &	L &	N	&	\nodata	&	N	&	N	&	\\ 
117	& Y	&	Y &	S &	Y	&	\nodata	&	Y	&	Y	&	\\ 
118	& Y	&	Y &	S?&	Y	&	N	&	Y	&	Y	&	Hatchell says starless (\# 82) with IRAC upper limits, but Enoch and c2d have a centered source.\\ 
119	& Y	&	Y &	S &	Y	&	Y	&	Y	&	Y	&	Two c2d sources.\\ 
120	& N	&	N &	L &	\nodata	&	\nodata	&	N	&	N	&	\\ 
121	& Y	&	Y &	S &	Y	&	Y	&	Y	&	Y	&	\\ 
122	& Y	&	N &	L &	\nodata	&	\nodata	&	N	&	N	&	\\ 
123	& Y	&	Y &	S &	Y	&	Y	&	Y	&	Y	&	\\ 
124	& Y	&	Y &	L &	N	&	\nodata	&	N	&	Y	&	\\ 
125	& Y	&	Y &	L &	N	&	\nodata	&	N	&	N	&	\\ 
126	& Y	&	Y &	S &	Y	&	Y	&	Y	&	Y	&	\\ 
127	& Y	&	Y &	L &	N	&	\nodata	&	N	&	N	&	\\ 
128	& Y	&	Y &	S?&	N	&	\nodata	&	Y	&	Y	&	Starless in Enoch, but close/faint 24$\micron$ source and c2d source.\\ 
129	& Y	&	N &	L &	\nodata	&	\nodata	&	N	&	N	&	\\ 
130	& N	&	N &	L &	\nodata	&	\nodata	&	N	&	N	&	\\ 
131	& N	&	N &	L &	\nodata	&	\nodata	&	N	&	N	&	\\ 
132	& Y	&	Y &	L &	N	&	\nodata	&	N	&	N	&	Outside of SCUBA.\\ 
133	& N	&	N &	L &	\nodata	&	\nodata	&	N	&	N	&	\\ 
134	& N	&	N &	L &	\nodata	&	\nodata	&	N	&	N	&	\\ 
135	& N	&	N &	L &	\nodata	&	\nodata	&	N	&	N	&	\\ 
136	& N	&	N &	L &	\nodata	&	\nodata	&	N	&	N	&	\\ 
137	& N	&	N &	L &	\nodata	&	\nodata	&	N	&	N	&	\\ 
138	& N	&	N &	L &	\nodata	&	\nodata	&	N	&	N	&	\\ 
139	& Y	&	N &	L &	\nodata	&	\nodata	&	N	&	N	&	Outside of SCUBA.\\ 
140	& N	&	N &	L &	\nodata	&	\nodata	&	N	&	N	&	\\ 
141	& Y	&	Y &	L &	N	&	\nodata	&	N	&	N	&	Outside of SCUBA.\\ 
142	& Y	&	Y &	L &	N	&	N	&	N	&	Y	&	Nearby 24$\micron$ source.\\ 
143	& N	&	Y &	S &	Y	&	Y	&	Y	&	Y	&	\\ 
144	& Y	&	Y &	L &	N	&	\nodata	&	N	&	N	&	\\ 
145	& Y	&	Y &	L &	N	&	N	&	N	&	Y	&	Diffuse 24$\micron$\\ 
146	& Y	&	Y &	L &	N	&	\nodata	&	N	&	Y	&	Diffuse 24$\micron$\\ 
147	& Y	&	Y &	L &	N	&	N	&	N	&	Y	&	Diffuse 24$\micron$\\ 
148	& Y	&	N &	L &	\nodata	&	\nodata	&	N	&	N	&	\\ 
149	& N	&	N &	S&	\nodata	&	\nodata	&	Y	&	Y	&	c2d source right at center.\\ 
150	& Y	&	Y &	L &	N	&	\nodata	&	N	&	Y	&	\\ 
151	& Y	&	Y &	L &	N	&	\nodata	&	N	&	Y	&	\\ 
152	& Y	&	Y &	L &	N	&	\nodata	&	N	&	Y	&	Unclear 24$\micron$ detection.\\ 
153	& Y	&	Y &	L &	N	&	\nodata	&	N	&	Y	&	Unclear 24$\micron$ detection.\\ 
154	& Y	&	Y &	L &	N	&	\nodata	&	N	&	Y	&	Unclear 24$\micron$ detection.\\ 
155	& N	&	N &	L &	\nodata	&	\nodata	&	N	&	Y	&	Diffuse 24$\micron$\\ 
156	& Y	&	Y &	L &	N	&	N	&	N	&	Y	&	\\ 
157	& Y	&	Y &	L &	N	&	N	&	N	&	Y	&	\\ 
158	& Y	&	Y &	L &	N	&	\nodata	&	N	&	Y	&	\\ 
159	& Y	&	Y &	L &	N	&	\nodata	&	N	&	Y	&	\\ 
160	& Y	&	Y &	S &	Y	&	Y	&	Y	&	Y	&	\\ 
161	& Y	&	Y &	S &	Y	&	Y	&	Y	&	Y	&	\\ 
162	& Y	&	Y &	S &	Y	&	Y	&	Y	&	Y	&	\\ 
163	& Y	&	Y &	L &	N	&	N	&	N	&	Y	&	Unclear 24$\micron$ detection.\\ 
164	& Y	&	Y &	S?&	Y	&	N	&	Y	&	Y	&	Enoch has a source in NH3SRC 164-165-169 cluster and two other c2d sources within this beam.\\ 
165	& Y	&	Y &	L? &	N	&	N	&	Y	&	Y	&	Enoch source is within 164, although at edge of beam.\\ 
166	& Y	&	Y &	L? &	Y	&	\nodata	&	N	&	N	&	Enoch says protostellar, but others methods say starless, and Bolocam core is very diffuse.\\ 
167	& Y	&	N &	L &	\nodata	&	\nodata	&	N	&	N	&	\\ 
168	& Y	&	Y &	S &	Y	&	\nodata	&	Y	&	Y	&	\\ 
169	& Y	&	Y &	L?&	Y	&	N	&	N	&	Y	&	This Enoch source is in NH3SRC 164.\\ 
170	& Y	&	Y &	L &	N	&	N	&	N	&	N	&	\\ 
171	& Y	&	Y &	L?&	N	&	Y	&	N	&	N	&	Hatchell says protostellar, but everything else says no.\\ 
172	& N	&	N &	L &	\nodata	&	\nodata	&	N	&	Y	&	\\ 
173	& Y	&	Y &	S &	Y	&	\nodata	&	Y	&	Y	&	\\ 
174	& Y	&	Y &	L &	N	&	N	&	N	&	Y	&	\\ 
175	& N	&	N &	L &	\nodata	&	\nodata	&	N	&	Y	&	\\ 
176	& Y	&	Y &	L &	N	&	N	&	N	&	Y	&	\\ 
177	& N	&	N &	S&	\nodata	&	\nodata	&	Y	&	Y	&	Two c2d sources within this core.\\ 
178	& Y	&	Y &	S &	Y	&	Y	&	Y	&	Y	&	\\ 
179	& N	&	N &	L &	\nodata	&	\nodata	&	N	&	Y	&	\\ 
180	& Y	&	Y &	L &	N	&	N	&	N	&	Y	&	\\ 
181	& Y	&	Y &	L &	N	&	\nodata	&	N	&	N	&	\\ 
182	& Y	&	N &	L &	\nodata	&	\nodata	&	N	&	N	&	\\ 
183	& Y	&	Y &	L &	N	&	N	&	N	&	Y	&	\\ 
184	& N	&	Y &	S &	Y	&	\nodata	&	Y	&	Y	&	c2d source at center.\\ 
185	& N	&	N &	L &	\nodata	&	\nodata	&	N	&	N	&	Outside of SCUBA.\\ 
186	& N	&	N &	L &	\nodata	&	\nodata	&	N	&	Y	&	Outside of SCUBA. Pointing misses star in tail of B5\\ 
187	& N	&	N &	L &	\nodata	&	\nodata	&	N	&	N	&	Outside of SCUBA.\\ 
188	& Y	&	Y &	L &	N	&	N	&	N	&	Y	&	\\ 
189	& Y	&	Y &	L?&	Y	&	N	&	N	&	Y	&	Hatchell defines a starless core. Larger Bolocam core includes NH3SRC 192.\\ 
190	& Y	&	N &	L &	\nodata	&	\nodata	&	N	&	N	&	\\ 
191	& Y	&	N &	L &	\nodata	&	\nodata	&	N	&	Y	&	\\ 
192	& Y	&	Y &	S &	Y	&	Y	&	Y	&	Y	&	\\ 
193	& N	&	N &	L &	\nodata	&	\nodata	&	N	&	N	&	Outside of SCUBA.\\
\tablenotetext{a}{These reference numbers are from \citet{Rosolowsky:2008}, which can be consulted for individual core positions and properties}
\tablenotetext{b}{L = Starless, S = Protostellar, L? = Uncertain Starless, S? = Uncertain Protostellar}
\tablenotetext{c}{\citet{Enoch:2008}}
\tablenotetext{d}{\citet{Hatchell:2007}}
\tablenotetext{e}{\citet{Evans:2007}, considered associated if a YSO candidate falls within 1 FWHM of the GBT beam (31\arcsec) of the pointing location.}
\tablenotetext{f}{\citet{Jijina:1999}}
\enddata
\label{IDs}
\end{deluxetable}

\clearpage

\section{Appendix B: Sensitivity to Random and Systematic Temperature Errors}
\label{AppendixB}

Many of the distributions we measure and describe depend critically on a temperature derived from modeling the NH$_3$ spectra in order to convert observed quantities into physical units. This dependence on temperature can be divided into two cases:

1) The distributions which use only the NH$_3$ and CCS data (T$_\mathrm{ex}$, T$_\mathrm{kin}$, and both non-thermal linewidths) require that our determination of temperature accurately measures the temperature of the bulk (H$_2$) gas. The most serious error possible is a systematic one. 

2) The distributions which also rely on the Bolocam maps (N(H$_2$), M$_\mathrm{dust}$, $\alpha$, X(CCS) and X(NH$_3$)) require that T$_\mathrm{kin}$ from NH$_3$ adequately approximates T$_\mathrm{dust}$. This is an estimate and may be biased in some way. We show in Appendix B that the best estimate for this discrepancy has a small influence on our results. Nonetheless, for this half of the analysis the probabilities quoted are valid only under the assumption that T$_\mathrm{kin}$ for NH$_3$ adequately approximates T$_\mathrm{dust}$.

\begin{figure}
\plottwo{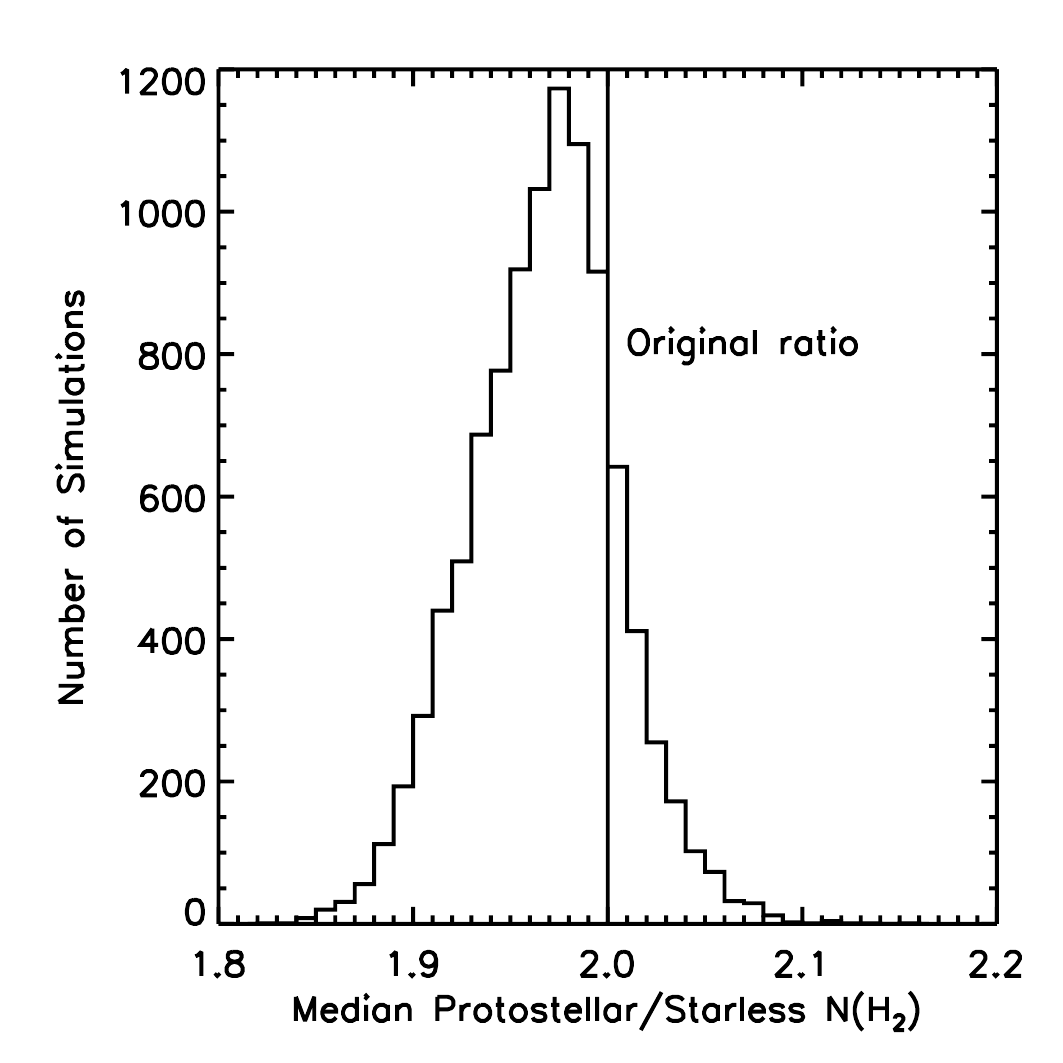}{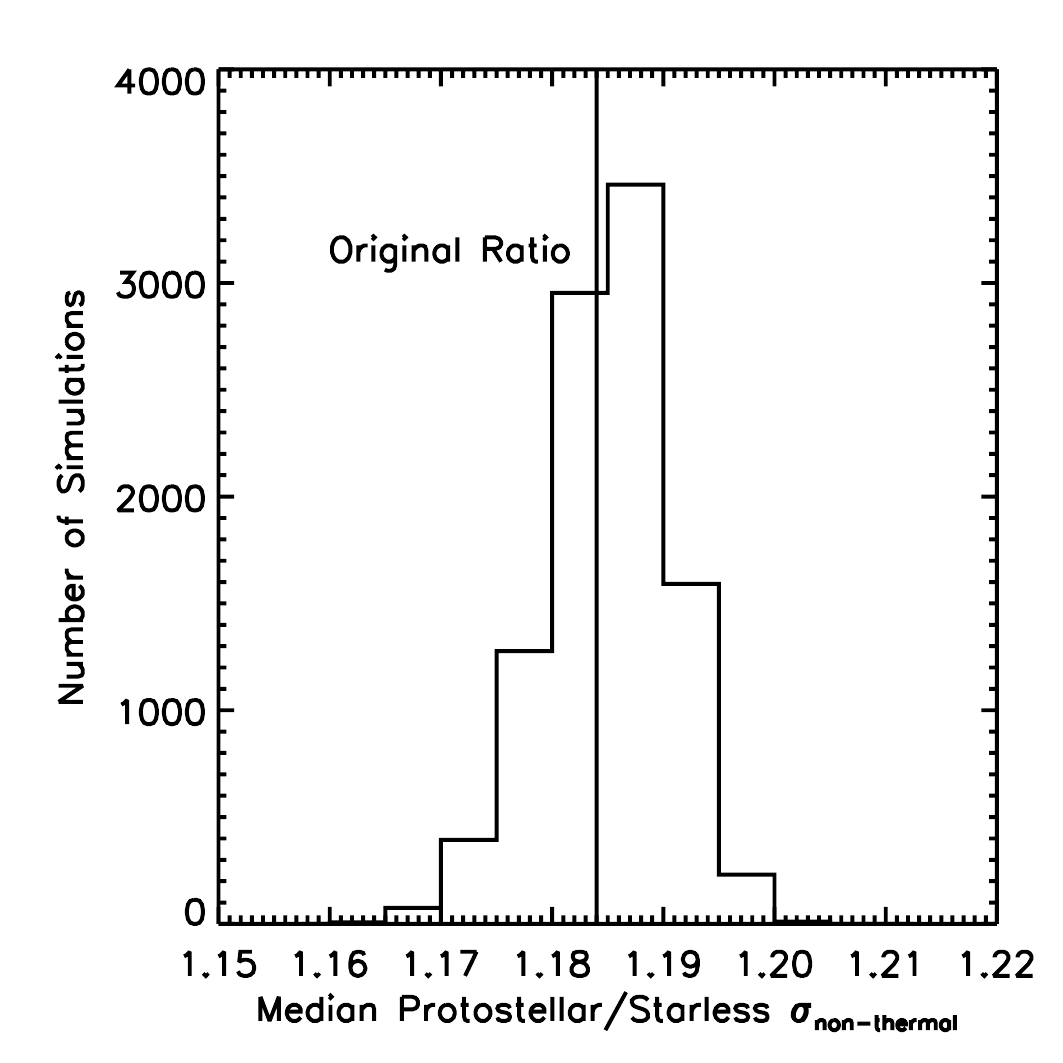}
\caption{Results of 10,000 simulations adding estimated random uncertainties to T$_\mathrm{kin}$ for N(H$_2$) (left panel) and for the non-thermal NH$_3$ linewidth (right panel). In both cases the median ratio of the protostellar and starless populations are shown. On the left, we see asymmetric scatter around the original value of 2, but no point approaches 1 (which would be the case where our statement that protostellar cores are at higher column density would be incorrect). On the right, a much smaller relative variation is present in non-thermal linewidths as this variable is less sensitive to changes in T$_\mathrm{kin}$. The original value of this ratio is 1.184.}
\label{Testa}
\end{figure}

\begin{figure}
\plotone{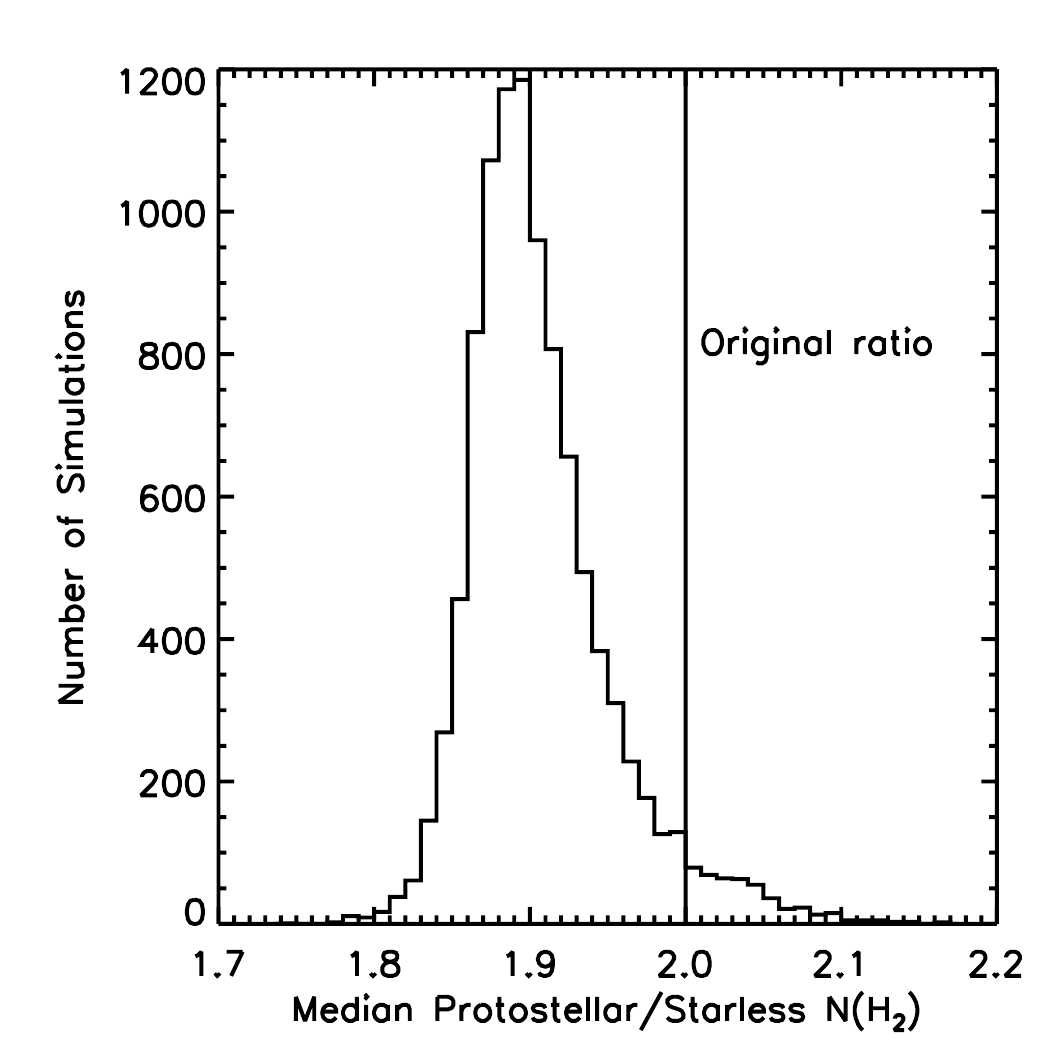}
\caption{Results of 10,000 simulations adding estimated random uncertainties to T$_\mathrm{kin}$ for N(H$_2$) and a 1 K underestimate of the dust temperature. The median ratio of the protostellar and starless populations is shown. The bias shifts the ratio from the original ratio of 2 down to peak near 1.9  with some asymmetric scatter. No point approaches 1 (which would be the case where our statement that protostellar cores are at higher column density would be incorrect).}
\label{Testb}
\end{figure}

\begin{figure}
\plotone{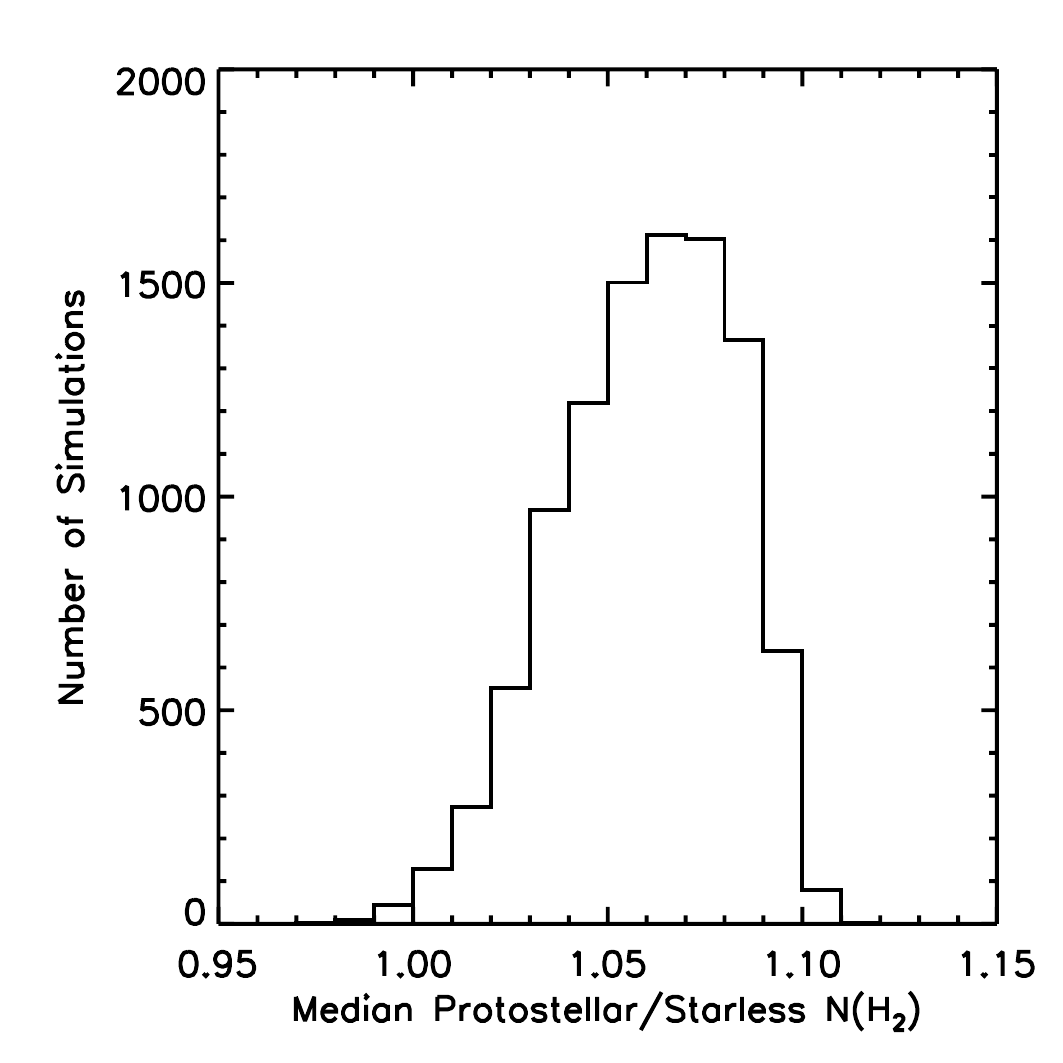}
\caption{Results of 10,000 simulations adding estimated random uncertainties to T$_\mathrm{kin}$ for N(H$_2$) and the assumption that NH$_3$ overestimates the dust temperature of starless cores by 2 K and underestimate the dust temperature by 2 K in protostellar cores. The median ratio of the protostellar and starless populations is shown. The original ratio of 2 has been shifted substantially, and this distribution now approaches 1. For larger systematic errors of this particular kind, the distinction between the populations is erased or reversed.} 
\label{Testc}
\end{figure}

\begin{figure}
\plottwo{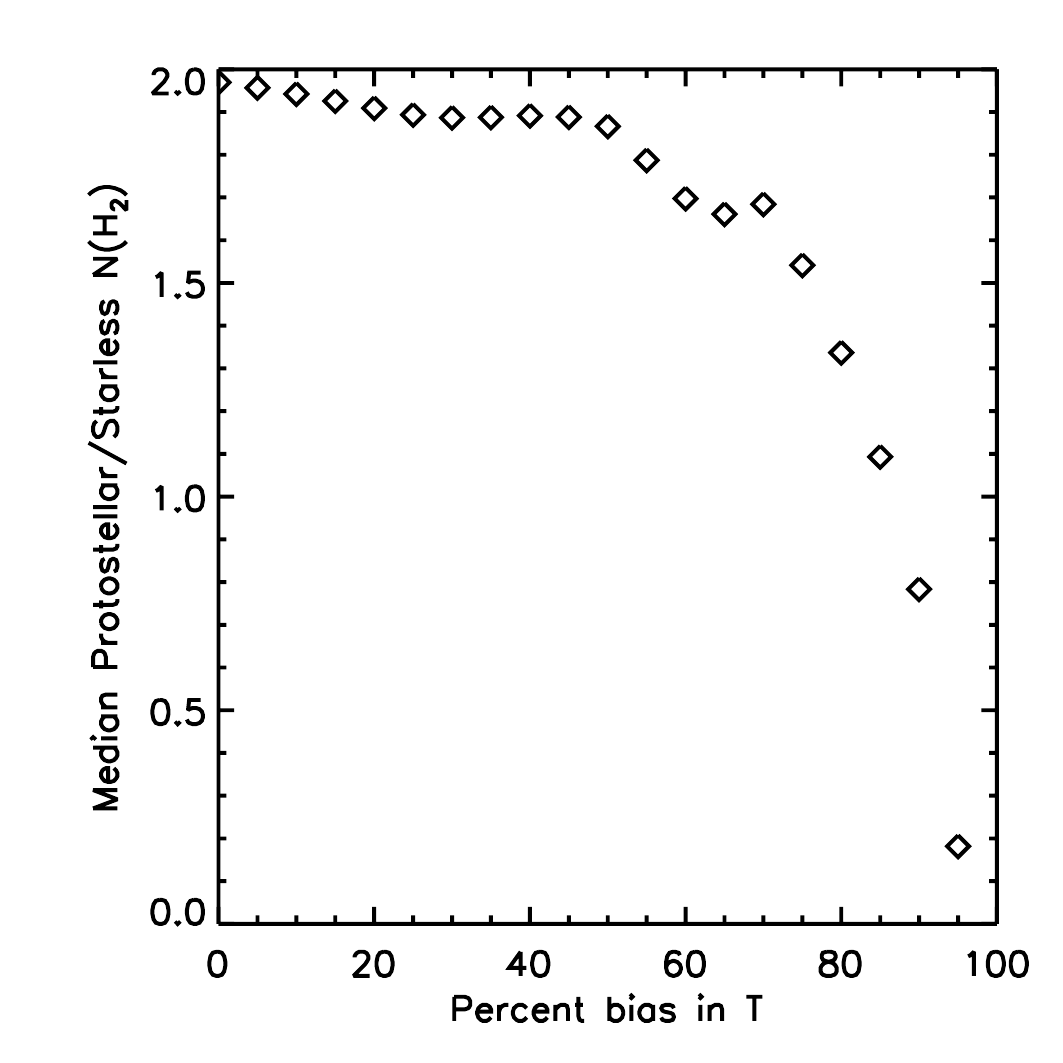}{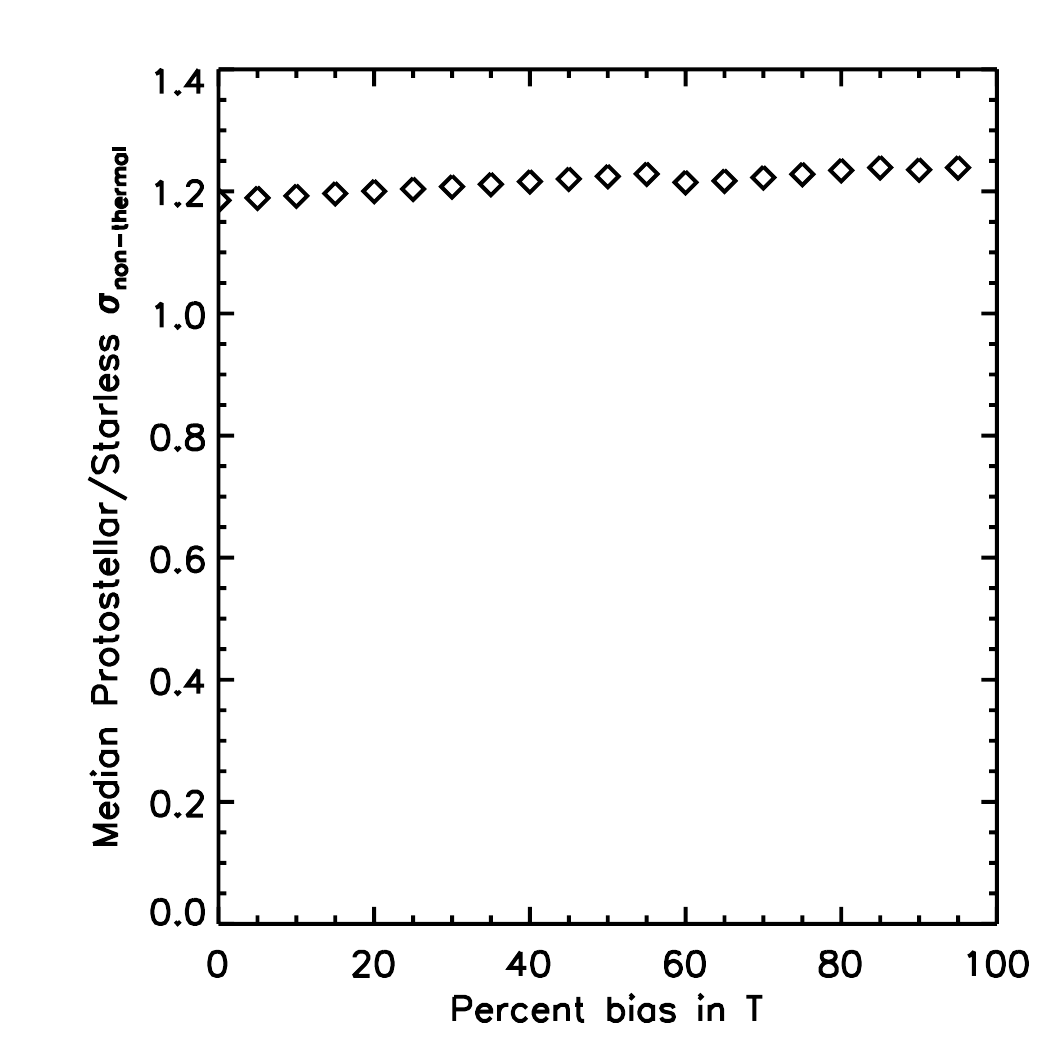}
\caption{The ratio of N(H$_2$) and non-thermal linewidth for our two populations under the assumption of a systematic problem in using NH$_3$ to determine the bulk gas temperature of the form T$_\mathrm{kin} = $ (1+$\delta$) T$_\mathrm{kin}$. In both cases the median ratio of the protostellar and starless populations are shown. For N(H$_2$) this systematic error needs to reach 60\% before significantly influencing the ratio, and the ratio does not reach equality until 80\%. The result for non-thermal linewidths are almost entirely robust to this sort of error.}
\label{Testd}
\end{figure}

As the two simplest quantities to assess, we perform experiments on the non-thermal linewidth of NH$_3$ (case 1) and the total column density (case 2). We perform 4 distinct Monte Carlo experiments:

a) We apply the random uncertainties of T$_\mathrm{kin}$ derived in \citet{Rosolowsky:2008} to both case 1 and case 2 for 10,000 individual realizations and calculate the percentage of cases in which our distinctions hold. The test we perform on the result is to simply take the median of the ratios for the protostellar and starless populations, discarding the cluster information for simplicity. Furthermore, the confidence intervals and P-values derived from ANOVA are not calculated using measurement errors (instead, they look at the width of the sample distributions). Thus, we do not expect our simulations to fully reproduce the confidence intervals present in ANOVA. As shown in Figure~\ref{Testa}, the addition of these random errors has a negligible influence on our results.

b) We apply our best estimate of the systematic offset between T$_\mathrm{kin}$ and T$_\mathrm{dust}$ to case 2 (1 K underestimate of the dust temperature) and again run 10,000 realizations with random uncertainties. We take this estimate of 1 K from \citet{Galli:2002} Figure 3, which shows the dust and gas temperatures within a starless core of the appropriate density range to be traced with Bolocam and NH$_3$. This figure is for the the low radiation case, which we believe to be appropriate for the relatively low intensity star-formation present in Perseus. As shown in Figure~\ref{Testb}, this sort of systematic offset has a small influence on our result.

c) We apply a pessimistic estimate of the systematic offset between T$_\mathrm{kin}$ and T$_\mathrm{dust}$ to case 2. Here, NH$_3$ over-estimates the temperature of starless cores by 2 K and under-estimates the temperature of protostellar cores by 2 K. This will increase the temperature differences between starless and protostellar cores, approximating the temperatures difference assumed in \citet{Enoch:2008} and \citet{Hatchell:2007}. We again run 10,000 realizations with random uncertainties. As shown in Figure~\ref{Testc}, such an offset severely diminishes (but does not eliminate) the difference between column densities towards protostellar and starless cores.

d) We adopt a systematic error such that T$_\mathrm{kin}$ in warm sources is underestimated by an arbitrarily chosen amount (New T$_\mathrm{kin} = $ (1+$\delta$) T$_\mathrm{kin}$). We allow $\delta$ for vary from 0 to 1 and examine at what value of $\delta$ the difference in these distributions is eliminated. As shown in  Figure~\ref{Testd}, significant bias is required to eliminate the difference in column density, and the difference in non-thermal linewidths is robust against an error of this sort.

These tests show that our results are very robust against the influence of random errors on the temperature. This is mostly due to the high-quality of the spectra, which allows a very precise estimate of the temperature. The non-thermal linewidth result is robust against systematic errors in the temperature determination. Systematic errors in the temperature determination can be constructed which eliminate the distinctions we see in quantities which rely on the dust emission map, particularly if these systematic differences work in different directions for the two populations (i.e. protostellar/starless). However, under our best estimate of the possible systematic error (Trial b) our results are affected by only a small amount. 

\end{document}